\documentclass[a4paper,10pt]{article}
\usepackage[utf8]{inputenc}
\usepackage[english]{babel}
\usepackage{csquotes}
\usepackage{cite}
\usepackage{authblk}
\usepackage{amsmath,amssymb,amsfonts}
\usepackage{algorithmic}
\usepackage{graphicx}
\usepackage{textcomp}
\usepackage{bmpsize}
\usepackage{xcolor}
\usepackage{setspace}
\usepackage{url}
\usepackage{xspace}
\usepackage{amsthm}
\usepackage{tablefootnote}
\usepackage{fullpage}
\usepackage{booktabs}
\usepackage[iso,english]{isodate}
\usepackage{tabularx, makecell, framed, enumitem}
\usepackage[nottoc,notlot,notlof]{tocbibind}
\usepackage{multicol}
\usepackage{beton}
\usepackage{euler}
\usepackage[frozencache=true,cachedir=_minted-eth2sec]{minted} 
\usepackage{longtable}
\usepackage{adjustbox}
\usepackage{newunicodechar}

\AtBeginDocument
{\DeclareFontShape\encodingdefault{ccr}{bx}{n}{<->sub*cmss/sbc/n}{}%
\DeclareFontShape\encodingdefault{ccr}{bx}{it}{<->sub*cmss/sbc/it}{}%
\DeclareFontShape\encodingdefault{ccr}{bx}{sl}{<->sub*cmss/sbc/sl}{}%
\DeclareFontShape\encodingdefault{ccr}{bx}{sc}{<->sub*cmss/sbc/sc}{}}

\newcommand\Warning{%
\makebox[1.4em][c]{%
\makebox[0pt][c]{\raisebox{.1em}{\small!}}%
\makebox[0pt][c]{\color{red}\Large$\bigtriangleup$}}}%
\newunicodechar{⚠}{\Warning}

\definecolor{ethblu}{RGB}{31,0,189}

\usepackage[colorlinks=true,allcolors=ethblu]{hyperref}
\usepackage{enumitem}
\setlist[itemize]{noitemsep}

\AtBeginDocument
{\DeclareFontShape\encodingdefault{ccr}{bx}{n}{<->sub*cmss/sbc/n}{}%
\DeclareFontShape\encodingdefault{ccr}{bx}{it}{<->sub*cmss/sbc/it}{}%
\DeclareFontShape\encodingdefault{ccr}{bx}{sl}{<->sub*cmss/sbc/sl}{}%
\DeclareFontShape\encodingdefault{ccr}{bx}{sc}{<->sub*cmss/sbc/sc}{}}

\usepackage[n,advantage,operators,sets,adversary,landau,probability,notions,logic,ff,mm,primitives,events,complexity,asymptotics,keys]{cryptocode}

\newcommand{\code}[1]{\texttt{#1}}
\renewcommand{\path}[1]{\textit{#1}}
\newcommand{\SK}{\mathsf{SK}}
\newcommand{\PK}{\mathsf{PK}}
\newcommand{\HtC}{\mathsf{H2C}}
\newcommand{\ECDH}{\mathsf{DH}}

\isodash{}
\date{\isodate\today}

\begin{document}

\vfill


\begin{center}
\includegraphics[width=7cm]{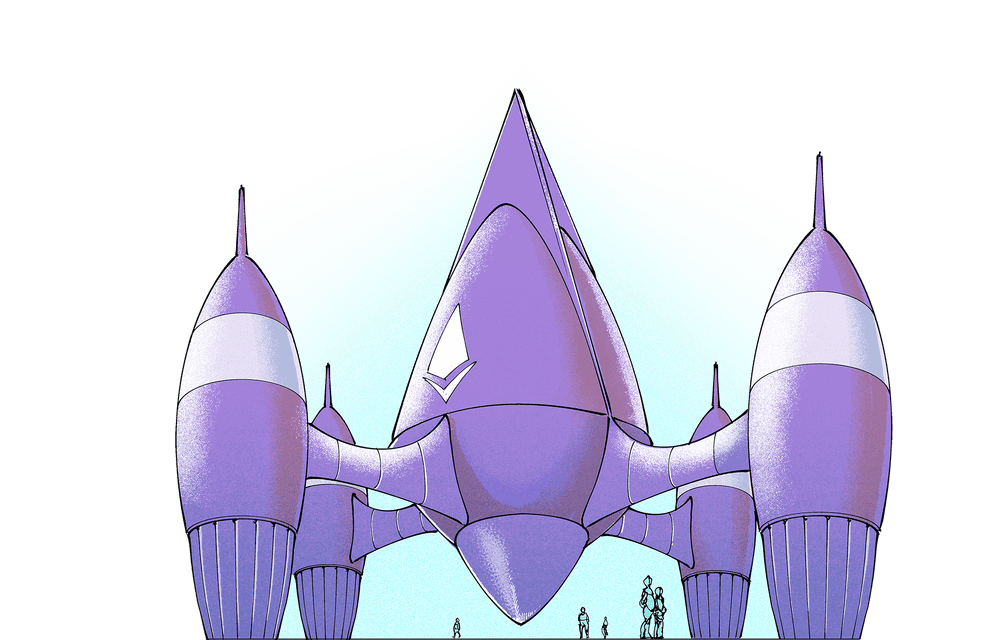}

\thispagestyle{empty}

\vspace{1cm}

{\LARGE
\sf \bf
Security Review of Ethereum Beacon Clients
}

\vspace{1cm}

JP Aumasson -- Taurus, Switzerland -- \href{mailto:jp@taurusgroup.ch}{jp@taurusgroup.ch}, \\ 
Denis Kolegov -- Tomsk State University, Russia -- \href{mailto:d.n.kolegov@gmail.com}{d.n.kolegov@gmail.com} \\
Evangelia Stathopoulou -- University College London, UK -- \href{mailto:evangelia.stathopoulou.20@ucl.ac.uk}{evangelia.stathopoulou.20@ucl.ac.uk}

\vspace{0.5cm}

Version \today

\vspace{0.5cm}

\emph{Supported by the Ethereum Foundation.}

\end{center}

\bigskip

\begin{abstract}
    The beacon chain is the backbone of the Ethereum's evolution
    towards a proof-of-stake-based scalable network.
    \textbf{Beacon clients} are the applications implementing the services
    required to operate the beacon chain, namely validators, beacon
    nodes, and slashers.
    Security defects in beacon clients could lead to loss of funds,
    consensus rules violation, network congestion, and other
    inconveniences.

    We reported more than \textbf{35 issues} to the beacon client
    developers, including various security improvements, specification
    inconsistencies, missing security checks, exposure to known
    vulnerabilities. None of our findings appears to be high-severity.
    We covered the four main beacon clients, namely Lighthouse (Rust),
    Nimbus (Nim), Prysm (Go), and Teku (Java).

    We looked for bugs in the logic and implementation of the new
    security-critical components (BLS signatures, slashing, networking
    protocols, and API) over a 3-month project that followed a
    preliminary analysis of BLS signatures code.
    We focused on Lighthouse and Prysm, the most popular clients, and
    thus the highest-value targets.
    Furthermore, we identify protocol-level issues, including replay
    attacks and incomplete forward secrecy.

    In addition, we reviewed the \textbf{network fingerprints} of beacon
    clients, discussing the information obtainable from passive and
    active searches, and we analyzed the \textbf{supply chain risk}
    related to third-party dependencies, providing indicators and
    recommendations to reduce the risk of backdoors and unpatchable
    vulnerabilities.

    Our results suggest that despite intense scrutiny by security
    auditors and independent researchers, the complexity and constant
    evolution of a platform like Ethereum requires regular expert
    review and thorough SSDLC practices.
\end{abstract}

\vspace{1cm}

\begin{center}

\raisebox{4mm}{\includegraphics[width=6.5cm]{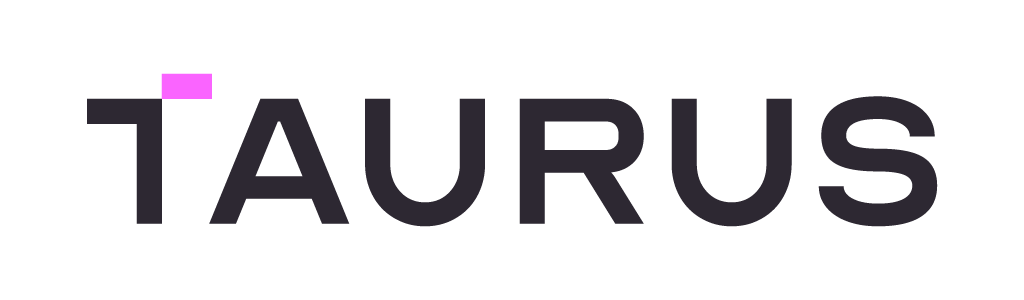}}
\quad
\quad
\includegraphics[width=2cm]{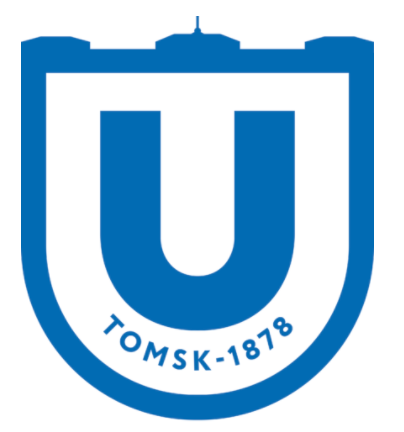}
\quad
\quad
\quad
\raisebox{5.5mm}{\includegraphics[width=4cm]{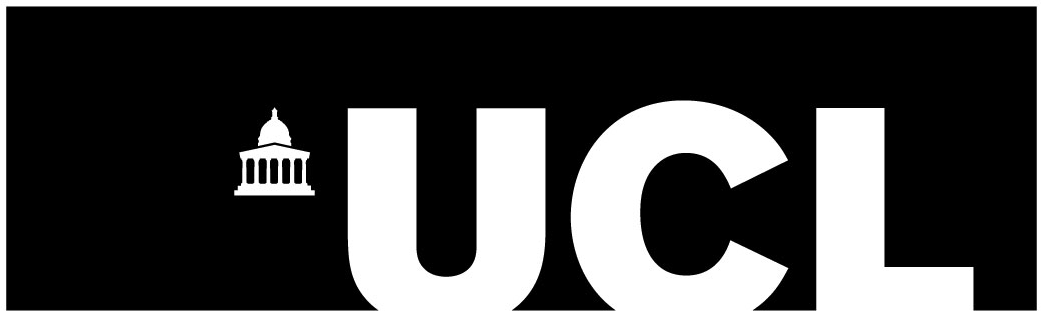}}
\hspace{1cm}
\,
\end{center}

\newpage

\newpage

\setcounter{tocdepth}{3}
\tableofcontents

\newpage

\section{Introduction}\label{sec:intro}

\subsection{Ethereum's Beacon Chain}

The \emph{beacon chain} is a chain separate from the main Ethereum
blockchain that acts as the governance platform behind ``Ethereum~2.0''
new mechanisms, including proof-of-stake-based consensus and sharding.
Software applications used to manage and interact with the beacon chain, sometimes just called ``Ethereum~2.0 clients'', are composed of two main components, as of Phase 0:

\begin{itemize}
    \item A \textbf{beacon node} service, which maintains a view of the beacon chain from the genesis block, manages validators, contributes randomness for validator assignment.
    \item A \textbf{validator} service, which proposes blocks and signs
    attestations for blocks proposed by other validators. Validators work with a beacon node to receive chain state information and propagate changes to other beacon nodes. Running a mainnet validator requires the staking of 32 ETH, locked via a deposit smart contract.
\end{itemize}
New security components introduced with the beacon chain notably include:
\begin{itemize}
    \item \textbf{BLS signatures}, the pairing-based, aggregation-friendly signatures used by validators to sign block attestations;
    \item \textbf{Slashing}, the punishing mechanism to prevent malicious behavior, implemented by certain beacon nodes (``slashers''), which submit slashing evidence to one or more validators for inclusion in an attested block.
\end{itemize}
These components and other new Ethereum features are extensively documented by the
Ethereum
project\footnote{\url{https://docs.ethhub.io/ethereum-roadmap/ethereum-2.0/eth-2.0-client-architecture/}}.

\subsection{This Project}

In March 2021, the Ethereum Foundation (EF) issued a ``Beacon chain security+testing RFP''\footnote{\url{https://notes.ethereum.org/@lsankar/security-rfp}}, calling for `` proposals that further the security and robustness of Ethereum’s beacon chain and the upcoming merge''.
Having followed the Ethereum developments, including BLS signatures implementations, we submitted a grant proposal for further in-depth security review of the beacon chain's critical components, covering (in rough order of priority)
\begin{itemize}
    \item BLS signatures
    \item Slashing 
    \item Peer-to-peer connections
    \item Beacon chain API 
\end{itemize}

We covered the four clients listed as targets in the EF RFP, namely Lighthouse, Nimbus, Prysm, and Teku, with as agreed with the EF a greater focus on Lighthouse and Prysm, as these appear to be the most used clients and the ones on which depend the greater amount of ETH assets (and consequently, the ones on which attackers are likely to invest the most resources).

We also covered the main implementation of BLS signatures, and its bindings, namely blst.
Both blst, Lighthouse, and Prysm have received intense security scrutiny
and testing, from their development teams, from the Ethereum
community, from security auditing firms, as well as from independent
researchers.
For example, a number of bugs were found using a dedicated differential
fuzzing framework, based on Vranken's initial
work\footnote{\url{https://github.com/sigp/beacon-fuzz}, \url{https://github.com/guidovranken/eth2.0-fuzzing}}.

\subsection{Our Results}\label{ap:issues}

Table~\ref{tab:issues} lists the GitHub issues that we opened to
notify project maintainers of potential security risks, in beacon
clients, BLS back-end, other dependencies, or specifications.
We excluded issues found to be invalid but listed some that, although
they did not lead to a patch, include interesting discussions about the
design choices.  We also included issues in beacon clients other than
the four ones that were the focus of our project.

Furthermore, we report a number of points that from our perspective need
more work and/or improvement to reduce the risk to an acceptable level.
These include cryptographically weak protocols (in~\S\ref{ssec:ndp},
~\S\ref{vcapi}) and risks from vulnerable and outdated dependencies
(in~\S\ref{sec:deps:res}).

\paragraph{Disclaimer.}
Given the uneven amount of time spent on the respective projects, it
makes little sense to draw conclusions from the relative number of
issues in each client---if we found more issues in
client A than on client B, it does not necessarily mean that A is less
secure; we may have spent much more time on A than on B, we might have
been more familiar with A's software stack, etc.

\begin{table}
    \centering
{
\footnotesize
\begin{tabular}{lll}
 \toprule
 \textbf{Issue Description} & \textbf{Component} & \textbf{Status} \\

 \midrule
 \multicolumn{3}{c}{\textbf{\href{https://github.com/cfrg/draft-irtf-cfrg-bls-signature/}{Specifications}}} \\
 \midrule
 \href{https://github.com/cfrg/draft-irtf-cfrg-bls-signature/issues/42}{Bit security level $<$ 128} & BLS specs & Fixed \\
 \href{https://github.com/cfrg/draft-irtf-cfrg-bls-signature/pull/41}{BLS parameters section number fix} & BLS specs & Fixed \\

 \midrule
 \midrule
 \multicolumn{3}{c}{\textbf{\href{https://github.com/supranational/blst}{supranational/blst}}} \\
 \midrule

\href{https://github.com/supranational/blst/commit/91f45b6be29ba2d5cc78d95f1642e0c20409c074}{Enforce limitation on IKM length} & BLS & Fixed \\

\midrule
\multicolumn{3}{c}{\textbf{\href{https://github.com/sigp/milagro_bls}{sigp/milagro\_bls}}} \\
\midrule

\href{https://github.com/sigp/milagro\_bls/issues/45}{Check that IKM is more than 32B in KeyGen} & BLS & Open \\

\midrule
\multicolumn{3}{c}{\textbf{\href{https://github.com/ChainSafe/bls/}{ChainSafe/bls}}} \\
\midrule

\href{https://github.com/ChainSafe/bls/issues/96}{BLS secret key validation is missing}& BLS & Confirmed \\

\midrule
\multicolumn{3}{c}{\textbf{\href{https://github.com/ChainSafe/blst-ts/}{ChainSafe/blst-ts}}} \\
\midrule

\href{https://github.com/ChainSafe/blst-ts/issues/24}{Incomplete key validation}& BLS & Fixed\\
\href{https://github.com/ChainSafe/blst-ts/issues/43}{Incorrect result for zero lengths arrays in aggregateVerify}& BLS & Fixed \\
\href{https://github.com/ChainSafe/blst-ts/issues/45}{Detect unsafe coefficients in verifyMultipleAggregateSignatures} & BLS & Fixed \\

\midrule
\midrule
\multicolumn{3}{c}{\textbf{\href{https://github.com/sigp/lighthouse}{sigp/lighthouse}}} \\
\midrule

\href{https://github.com/sigp/lighthouse/issues/2102}{Missing check on seed and password length}& BLS & Open\\
\href{https://github.com/sigp/lighthouse/issues/2438}{API token can be read from a log file by any user}& API & Fixed \\
\href{https://github.com/sigp/lighthouse/issues/2437}{File permissions for validator client API keys are insecure}& API & Fixed\\
\href{https://github.com/sigp/lighthouse/issues/2468}{Possible DoS via /eth/v1/validator/duties/attester}& API & Fixed\\
\href{https://github.com/sigp/lighthouse/issues/2512}{VC: Requests may not contain Authorization header with API token} & API & Fixed \\
\href{https://github.com/sigp/lighthouse/issues/2511}{VC: Response headers are not signed} & API & Confirmed \\

\midrule
\multicolumn{3}{c}{\textbf{\href{https://github.com/status-im/nimbus-eth2}{status-im/nimbus-eth2}}} \\
\midrule

\href{https://github.com/status-im/nimbus-eth2/issues/2693}{Insufficient
private key validation (covers nim-blscurve)} -- also reported in
\href{https://github.com/status-im/nim-blscurve/issues/114}{nim-blscurve}& BLS &
Fixed\\

\href{https://github.com/status-im/nimbus-eth2/issues/2695}{Missing pubkey and signature validation for blsVerify()?}& BLS & Fixed\\
\href{https://github.com/status-im/nimbus-eth2/issues/2734}{Possible DoS via beacon node API endpoints if the API exposed to untrusted parties}& API & Open \\

\midrule
\multicolumn{3}{c}{\textbf{\href{https://github.com/prysmaticlabs/prysm/}{prysmaticlabs/prysm}}} \\
\midrule

\href{https://github.com/prysmaticlabs/prysm/issues/8819}{Missing input validation in SecretKeyFromBigNum}& BLS & Fixed \\
\href{https://github.com/prysmaticlabs/prysm/issues/9098}{Detect unsafe coefficients in VerifyMultipleSignatures}& BLS & Fixed \\
\href{https://github.com/prysmaticlabs/prysm/issues/9091}{No length check in AggregatePublicKeys} & BLS & Fixed \\
\href{https://github.com/prysmaticlabs/prysm/issues/8863}{Update go-libp2p-noise to release v0.2.0}& Libp2p-noise & Fixed \\
\href{https://github.com/prysmaticlabs/prysm/issues/9015}{jwt-go library is vulnerable to CVE-2020-26160}& RPC & Fixed\\
\href{https://github.com/prysmaticlabs/prysm/issues/9357}{Use golang-jwt/jwt implementation of JWT} & RPC & Fixed \\
\href{https://github.com/prysmaticlabs/prysm/issues/9247}{Possible DoS via beacon node API endpoints if the API exposed to untrusted parties} & API & Fixed\\
\href{https://github.com/prysmaticlabs/prysm/issues/9293}{Add certificate-based client-side authentication} & API & Open \\
\href{https://github.com/prysmaticlabs/remote-signer/issues/14}{No TLS client authentication in gRPC} & API & Open \\
\href{https://github.com/prysmaticlabs/prysm-web-ui/issues/178}{Add HTTP Secure Headers}& UI & Open\\
\href{https://github.com/prysmaticlabs/prysm-web-ui/issues/187}{Logout endpoint doesn't require a valid JWT token} & UI & Open \\

\midrule
\multicolumn{3}{c}{\textbf{\href{https://github.com/ConsenSys/teku/}{ConsenSys/teku}}} \\
\midrule

\href{https://github.com/ConsenSys/teku/issues/4111}{Public key aggregation ambiguous infinite points handling}& BLS & Fixed \\
\href{https://github.com/ConsenSys/teku/issues/4112}{Detect unsafe coefficients in fast BLS verification}& BLS & Fixed \\
\href{https://github.com/ConsenSys/teku/issues/4025}{Incorrect BLS key validation}& BLS & Won't fix \\
\href{https://github.com/ConsenSys/teku/issues/4112}{Detect unsafe coefficients in fast BLS verification} & BLS & Fixed \\

\midrule
\multicolumn{3}{c}{\textbf{\href{https://github.com/libp2p/specs/tree/master/noise}{Libp2p-noise}}} \\
\midrule

\href{https://github.com/libp2p/specs/issues/355}{Static key signature does not depend on a peer's challenge}& Libp2p-noise spec & Confirmed \\

\midrule
\multicolumn{3}{c}{\textbf{\href{https://github.com/ChainSafe/lodestar}{ChainSafe/lodestar}}} \\
\midrule

\href{https://github.com/ChainSafe/lodestar/issues/2555}{No BLS public key validation due to validate parameter missing}& BLS & Open \\
\href{https://github.com/ChainSafe/lodestar/issues/2796}{Improper nonce handling in Noise handshake} & Libp2p-noise & Closed \\
\href{https://github.com/ChainSafe/js-libp2p-noise/issues/102}{Improper nonce handling} & Libp2p-noise & Open \\ 
\href{https://github.com/symbolicsoft/noiseexplorer/issues/1}{Improper nonce handling in Go} & NoiseExplorer & Fixed \\ 

\midrule
\multicolumn{3}{c}{\textbf{\href{https://github.com/ethereum/trinity}{ethereum/trinity}}} \\
\midrule

\href{https://github.com/ethereum/trinity/issues/2117}{Incomplete BLS key validation} & BLS & Won't fix\footnote{Because ``The Trinity
Ethereum client is no longer being maintained or developed''.} \\
 
\bottomrule
\end{tabular}
}
\caption{List of issues reported, including security improvements,
specification inconsistencies, missing security checks, exposure to
known vulnerabilities.}\label{tab:issues}
\end{table}

\section{Beacon Clients}\label{sec:sw}

Table~\ref{tab:clients} shows an overview of the four beacon clients
reviewed, and the sections below present further information  about
their security, their implementation of BLS and P2P protocols, public open
issues, and previous security audits.

\begin{table}[ht]
\begin{tabular}{p{40mm}lp{40mm}ll}
    \toprule
    \textbf{Client \& Repository} & \textbf{Language} & \textbf{Developers} & \textbf{Repo Stars} & \textbf{Open/Closed Issues} \\
    \midrule
    Lighthouse \newline \href{https://github.com/sigp/lighthouse/}{sigp/lighthouse} & Rust & Sigma Prime \newline \href{https://sigmaprime.io}{sigmaprime.io} & 1.3k & 100/846 \\  
    Nimbus \newline \href{https://github.com/status-im/nimbus-eth2}{status-im/nimbus-eth2} & Nim & Status \newline \href{https://status.im/}{status.im} & 222 & 152/526 \\
    Prysm \newline \href{https://github.com/prysmaticlabs/prysm/}{prysmaticlabs/prysm/} & Go & Prysmatic Labs \newline \href{https://prysmaticlabs.com}{prysmaticlabs.com} & 2.2k & 114/2016 \\
    Teku \newline \href{https://github.com/ConsenSys/teku}{ConsenSys/teku} & Java & ConsenSys \newline \href{https://www.consensys.net/}{consensys.net} & 281 & 82/1251 \\
    \bottomrule
\end{tabular}
\caption{Overview of the beacon clients reviewed, as of 20210913.}\label{tab:clients}
\end{table}

\subsection{Lighthouse}

Lighthouse's README describes it as ``Security-focused. Fuzzing techniques have been continuously applied and several external security reviews have been performed.''
Fuzzing is not performed in CI but ``All finalized crates are to go through a series of extensive fuzzing.''\cite{LhSC}.  
The initial fuzzing setup is described in a 2019 post\cite{LhFU}, notably using libFuzzer via \code{cargo-fuzz}.
The fact that \href{https://github.com/sigp/beacon-fuzz}{sigp/beacon-fuzz} has fewer trophies for Lighthouse than for any of the three other clients reviewed here may be seen as an indicator of Lighthouse's care for security.

The April 2021 Lighthouse Research Report\footnote{\url{https://drive.google.com/file/d/12flM_E_A3DldbOUe8EYEN6Bg2bi4Yl6w/view}} contains
a good introduction to the Lighthouse client design and implementation, and considers key challenges, engineering efforts, major optimizations,
development process peculiarities, and the roadmap.

\paragraph{BLS.}

Rust bindings of blst (part blst's distribution).

\paragraph{Networking.}

Own Rust implementations of libp2p and discv5, in \path{beacon\_note/eth2\_libp2p} and the repository \href{https://github.com/sigp/discv5}{sigp/discv5}. 

\paragraph{Open Issues.}

At the time of writing, open GitHub Issues with the \code{security} label included low-risk issues\footnote{\url{https://github.com/sigp/lighthouse/issues?q=is\%3Aopen+is\%3Aissue+label\%3Asecurity}} open between June and October 2020: ``Fork choice timing attack'', ``Out-of-date dependencies'', ``Parasitic voluntary exits '', ``Rethink and test fork handling in op pool''. 
Among the other open issues, the following two are the only ones directly related to security risks: ``Zeroize in BLST''\footnote{\url{https://github.com/sigp/lighthouse/issues/1908}} and ``Incorrect zeroizing of newtype structs''\footnote{\url{https://github.com/sigp/lighthouse/issues/1789
}}.

\paragraph{Audits.}

In June 2020 Trail of Bits completed a first security audit of Lighthouse, the report does not seem to have been published.
Sigma Prime commented~\cite{Lh26} that ``no critical issues were found''.
In October 2020 Lighthouse cited~\cite{Lh30} a second Trail of Bits audit round, as well as audit by NCC.
These covered all the critical features of Lighthouse, such as the p2p protocol, the API, the validation logic, and deserializations.

\subsection{Nimbus}

Nimbus targets resource-constrained platforms (citing Raspberry Pis and mobile devices) and is written in Nim, a language that
``provides memory safety by not performing pointer arithmetic, with optional checks, traced and untraced references and optional non-nullable types.''\footnote{\url{https://nim-lang.org/faq.html}}.

Nimbus offers an ``Auditors' book''\cite{Nhb}, which briefly describes its threat model, and provides useful information for security auditors, but is incomplete, having a number of sections left empty.
A ``Nimbus Eth2 Stack'' diagram\footnote{\url{https://miro.com/app/board/o9J_kvfytDI=/}} describes its architecture and high-level implementation.

\paragraph{BLS.}

\href{https://github.com/status-im/nim-blscurve}{status-im/nim-blscurve}, a dedicated Nim interface to BLS implementation back-ends.
This uses blst for x86\_64 and ARM64 architectures, and MIRACL\footnote{\url{https://github.com/miracl/core}} for other architectures (including ARM Cortex-M0/M4, ESP32, MIPS32, RISC-V).

\paragraph{Networking.}

Most of the networking logic is in another repository, \href{https://github.com/status-im/nim-eth}{status-im/nim-eth}, also maintained by Status.

\paragraph{Open Issues.}

At the time of writing, open GitHub Issues with the \code{security} included 9 improvement proposals\footnote{\url{https://github.com/status-im/nimbus-eth2/issues?q=is\%3Aopen+is\%3Aissue+label\%3Asecurity}}, such as reduced exposure of private keys, integration of Clang sanitizers, and a discussion about invalid BLS signatures of more general interest\footnote{\url{https://github.com/status-im/nimbus-eth2/issues/555}}.

\paragraph{Audits.}

In May 2020, the Nimbus maintainers (Status) issued an RFP for a security audit~\cite{NimRFP}, and a short summary was published in September~\cite{NimUp}, without sharing the report nor the audit team.

\subsection{Prysm}

Prysm integrates extensive fuzzing\footnote{\url{https://github.com/prysmaticlabs/prysm/tree/develop/fuzz}}, using the standard Go fuzzing toolchain, with libfuzzer as fuzzing engine, and notably covering block validation, RPC endpoints, SSZ decoding.
Fuzz tests are run in the GitLab CI via fuzzit. 

Prysm is the only beacon client to sport a graphical UI, as a web application for local configuration\footnote{\url{https://docs.prylabs.network/docs/prysm-usage/web-interface/}}.

\paragraph{BLS.}

Go bindings of blst (part of blst's distribution).

\paragraph{Networking.}

\href{github.com/libp2p/go-libp2p}{libp2p/go-libp2p}, and its own implementation of the discv5 logic.

\paragraph{Open Issues.}

The open security-related issues only include two about version updates and a meta-issue tracking issues reported by the two audits as well as proposed improvements\footnote{\url{https://github.com/prysmaticlabs/prysm/issues/7514}}.

\paragraph{Audits.}

In October 2020 Trail of Bits completed a first security audit of Prysm~\cite{ToBprysm}. 
The audit found no critical issues and one high-severity issue
related to a failure scenario causing a user's password to be logged.
In October 2020 Quantstamp also published an audit report~\cite{QsPrysm}, which contained 4 high risk issues (3 of them were fixed and 1 was acknowledged).

Both reports noted shortcomings in the SDLC process:
\begin{itemize}
  \item Many dependencies (including core components like bbolt and libp2p) were outdated and included known bugs;
  \item Many pieces of the code lack unit tests.
\end{itemize}

\subsection{Teku}

Teku is advertized as ``built to meet institutional needs and security requirements''\footnote{\url{https://consensys.net/knowledge-base/ethereum-2/teku/}} and includes basic fuzz test suites, which notably cover the slashing mechanism.

\paragraph{BLS.} Teku uses blst's Java bindings, with an additional wrapper layer in \path{tech/pegasys/teku/bls/impl}.

\paragraph{Networking.} Own Java implementation in \path{tech/pegasys/teku/networking/}.

\paragraph*{Open Issues.}

We did not find open issues that appeared directly security-related.

\paragraph*{Audits}

In October 2020, Quantstamp complete an audit of Teku, whose details were published~\cite{QsTeku}, with documentation of issues' resolution.

\section{BLS Signatures}\label{sec:bls}

What's known as ``BLS signatures'' encompasses the original 2001 Boneh-Lynn-Shacham pairing-based signatures~\cite{BLS01} and the 2018 collective signing extensions~\cite{BDN18,BDN18b} based on the 2003 work on aggregate signatures. 
Under this umbrella term, BLS signatures are:

\begin{itemize}
    \item Deterministic
    \item Non-interactive
    \item Short (one group element)
    \item Simple, given a pairing operator
    \item Easily adapted to support collective signing operations:
    \begin{itemize}
        \item Aggregation of signatures and public keys for $n$-of-$n$ signing
        \item Threshold signing ($t$-of-$n$) 
        \item Batch verification
    \end{itemize}
\end{itemize}
For a similar security level, in the single-signature setting, BLS signing
 is about as fast as with ECDSA or Schnorr signatures, but verification
 is slower because of the two pairings involved\footnote{See benchmarks
 in
 \url{https://www.mintlayer.org/news/2021-05-17-why-mintlayer-adopts-bls-signature/}}.
Efficient aggregation is the feature that drove Ethereum to choose BLS signatures and was described as ``a pragmatic medium-term solution to the signature verification bottleneck of sharding and Casper''~\cite{PragAg}.
Note that BLS signatures are not post-quantum.

We provide a high-level description of BLS signatures, where we tried to
use notations close to those in the IETF draft.  
We only describe the functional operation and \emph{do not include the
security checks}.

BLS signatures use a non-degenerate bilinear pairing\footnote{Although
BLS signatures were proposed in the paper titled ``Short Signatures from the Weil Pairing'', they now usually rely on the Ate pairing, discovered after that paper was published, and are more efficient than Weil's pairing.} $e:\GG_1\times\GG_2\to \GG$, where $\GG_1$ and $\GG_2$ are distinct isomorphic groups, thus of same prime order. 
The pairing satisfies
\begin{itemize}
    \item $e(P+Q,R)=e(P,R)e(Q,R)$
    \item $e(P,Q+R)=e(P,Q)e(P,R)$
\end{itemize}
which implies $e(nP,Q)=e(P,nQ)=e(P,Q)^n$ for a scalar $n$.

A key pair is (scalar $\SK$, point $\PK=\SK\times P$), where $P$ is a
generator of the curve's group, which can be $\GG_1$ or $\GG_2$, but the beacon chain uses $\GG_1$ so we'll stick to this convention (see details in~\S\ref{ssec:blsbeac}).

Signing a message $M$ consists in hashing the message to a curve point, denoted $\HtC:\{0,1\}^\star \to \GG_2$ and multiplying it with the secret key:
\[
    \SK \times \HtC(M) \in \GG_2
\]

Verifying a signature $S$ of a message $M$ then consists in computing
two pairings and checking their equality: $e(\PK, \HtC(M))$, and
$e(P,S)$, where $P$ is the generator of $\GG_1$ such that $\SK\times
P=\PK$.
Indeed, 
\[
e(\PK,\HtC(M))=e(\SK\times P,\HtC(M))=e(P,\HtC(M))^\SK =
e(P,\SK\times\HtC(M)).
\]

\subsection{Beacon Chain Integration}\label{ssec:blsbeac}

Ethereum adopted BLS signatures as specified in the IETF BLS signature
draft~\cite[Ap.A]{BLSIE}, which uses the BLS12-381 parameters specified
in~\cite[\S4.2.1]{PFC}, including the group $\GG_1$ and $\GG_2$ definition.
BLS12-381 is an instance of the Barreto--Lynn--Scott family BLS12~\cite{BLScurve} proposed by Zcash~\cite{BLS12381}\footnote{Whereas a 128-bit security level is usually expected of
BLS12-381 BLS signatures, it's strictly speaking allegedly lower than
120, as discussed in~\cite[\S3.2]{PFC} and~\cite[p8]{NCCBLS}.}.
But this is not alarming~\cite{TMC}.

The hash-to-curve algorithm is the one specified in the v09 of the Internet Draft ``Hashing to Elliptic Curves''~\cite{H2CIE} (based on~\cite{WB19,SBCDK08}).
Specifically, the variant used is
\code{BLS\_SIG\_BLS12381G2\_XMD:SHA-256\_SSWU\_RO\_POP\_}, which uses
the ``minimal-pubkey-size'' parameter, or the convention of using public
keys in $\GG_1$ and signatures in $\GG_2$.

The IETF draft not only describes the mathematical, functional operations but aims to be an implementable specification, including:
\begin{itemize}
    \item Security pre-condition verification (such as checking that points belong to the right subgroup and are not the identity).
    \item Encoding and decoding to/from bytes.
    \item Clear subroutines for each operation.
    \item Mitigation against rogue-key attacks via proofs of private key possession.
    \item Cipher suites definition.
    \item Test vectors (to appear in the final version).
\end{itemize}
This is a challenging and commendable effort, that will likely contribute to more consistent and safer implementations.
In practice, implementations may differ a bit, for example when
implementing the following \textsf{CoreVerify} algorithm:
\begin{verbatim}
    1. R = signature_to_point(signature)
    2. If R is INVALID, return INVALID
    3. If signature_subgroup_check(R) is INVALID, return INVALID
    4. If KeyValidate(PK) is INVALID, return INVALID
    5. xP = pubkey_to_point(PK)
    6. Q = hash_to_point(message)
    7. C1 = pairing(Q, xP)
    8. C2 = pairing(R, P)
    9. If C1 == C2, return VALID, else return INVALID
\end{verbatim}
This returns the same \textsf{INVALID} for all error types, whereas
implementations may return different error codes and/or messages.
We can imagine scenarios where returning the same \textsf{INVALID} error could facilitate certain attacks (fault injection with inaccurate faults), but in Ethereum's use case it's unlikely to be an issue. 

We reviewed that the security checks mandated by the IETF draft were
done correctly in each implementation (and reported a number of issues):
\begin{enumerate}
    \item ``IKM MUST be infeasible to guess''
    \item ``IKM MUST be at least 32 bytes long.'' 
    \item ``Implementations of the underlying pairing-friendly elliptic curve SHOULD run in constant time.'' (With respect to the key, not necessarily the message.)
    \item The security checks defined by \code{pubkey\_subgroup\_check()}, \code{signature\_subgroup\_check()}, \code{KeyValidate()}, and \code{PopVerify()} are properly implemented, used where they must be, and their return value processed correctly processed.
    \item \code{hash\_to\_point()} and \code{hash\_pubkey\_to\_point()} functions implemented using a secure hash-to-curve algorithm.
    \item Each implemented signature scheme is protected against
    rogue-key attacks~\cite{rogue}, with the exception, by design, of \code{FastAggregateVerify()}
    \item The secret key SK must be ``such that 1 $\leq$ SK $<$ r'', and thus enforced at generation and signing.
\end{enumerate}
Note that the second version of the IETF draft did not include the identity check in the key validation, only the subgroup check.
This may be a source of confusion if implementers refer to an older version of the document.

Depending on the usage of BLS signatures, additional security checks may
be needed.
For example, in a use case where distinct signers jointly issue a
signature, it may be valuable to check that all public keys are distinct.
Also, ``splitting-zero attacks''~\cite{zero} highlight a property that
an attacker can determine combinations of private keys such that the sum
aggregate will be zero\footnote{See also \url{https://github.com/cryptosubtlety/zero}.}.

\subsection{Aggregation and Rogue-Key Attacks}

The \textsf{Aggregate} function of the IETF draft adds up signatures into a single point, then \textsf{AggregateVerify} takes a signature (aggregated) $R$, $n$ public keys $\PK_i$ and as many messages $M_i$ and verifies that all messages are distinct, and computes the product
\[
    \prod_{i=1}^n e(\PK_i,\HtC(M_i)) = \prod_{i=1}^n e(\SK_i\times
    P,\HtC(M_i)) = \prod_{i}^n e(P, \SK_i \times \HtC(M_i)) = e(P,R)
\]
and verifies that it matches $e(P,R)$. 

When the same message is signed by all parties, the sequential computation of pairings can be replaced by additions and a single pairing. 
This is done in the \textsf{FastAggregateVerify} function, which must
come with proofs of possession of the public keys' private keys, to
thwart \emph{rogue-key attacks}.

Rogue-key attacks allow an attacker to forge aggregate signatures of the same message given one or more public keys of other signers. 
The idea is that given a public key $\PK_1$, an attacker can create $\PK_2=r \times P - \PK_1$, therefore $r \times \HtC(M)= (\SK_1+\SK_2)\times\HtC(M)$ will be a valid aggregate signature of $M$---with the caveat that the attacker doesn't know $\SK_2=r - \SK_1$.

This attack works when the aggregation includes duplicate messages. 
A mitigation is thus to ensure that all messages signed are distinct, but this eliminates large classes of important use cases.
Another mitigation is to force each user to prove that they know their private key, which is why the IETF draft includes proofs of possession (the routines \textsf{PopProve}, \textsf{PopVerify}), which are essentially single-signer signature and verification.
However, it may be unclear to readers that the ``fast'' verification routine \textsf{FastAggregateVerify} must be accompanied with some assurance, as a previous proof of possession, that signers know the private key.

Note that~\cite{BDN18} describes a trick to create an aggregation mechanism that does not require proofs of possession to be safe against rogue-key attacks, by multiplying each signature by the hash of its corresponding public key.
However, this variant is not specified in the IETF draft and is not used in Ethereum clients.

\subsection{Fast Batch Verification}

To verify multiple aggregate signatures efficiently, Buterin proposed an efficiency optimization~\cite{VBopt}, which is implemented by all the clients we reviewed in this paper. 
However, no formal specification or security analysis is available, and the IETF draft does not describe it.
This technique works as follows, given $n$ aggregate signatures $S_i$, each over $m_i$ messages:
\[
    e(S_i, P) = \prod_{j=1}^{m_i} e(P_{i,j}, M_{i,j}), i=1,\dots,n
\]
The naive method thus consists in checking these $n$ equalities, which involves $n+\sum_{i=1}^n m_i$ pairing operations.

One can further aggregate signatures to slightly reduce the number of pairings, as follows:
the verifier generates $n$ random values $1\leq r_i<n$, and aggregates all aggregate signatures into a single one:
\[
S^\star = r_1S_1 + \cdots r_n S_n
\]
the verifier also ``updates'' the signed messages (as their hashes to the curve) to integrate the coefficient of their batch, defining 
\[
    M_{i,j}'=r_i M_{i,j}, i=1,\dots,n, j=1,\dots,m_i
\]
Then verification can be done by checking
\[
e(S^\star,G)=\prod_{i=1}^n \prod_{j=1}^{m_i} e(P_{i,j},M_{i,j}')
\]
Verification thus saves $n-1$ pairing operations, but adds $n+\sum_{i=1}^n m_i$ scalar multiplications. 
However, if the verification fails then the verifier can't tell which (aggregate) signature is invalid.

The security goal of batch verification is that it should succeed if and
only if all signatures would be individually successfully verified, when
one or more (possibly all) of the signers may be maliciously colluding.
See~\cite{CHP12} for a rigorous treatment of batch verification security.

The post and thread~\cite{VBopt} include a basic analysis and discuss variants and optimizations.
It is easy to verify that this rewriting of the verification works, as
long as the $r_i$'s are in $[1,r)$ and that group elements have been
properly validated to be non-zero.
We noted that several of the beacon clients did not include this safety check, but they added it after we reported the problem.

As the post describes, random coefficients are required to prevent a trivial attack:
\begin{quote}
the randomizing factors are necessary: otherwise, an attacker could make a bad block where if the correct signatures are $C_1$ and $C_2$, the attacker sets the signatures to $C_1+D$ and $C_2-D$ for some deviation $D$. A full signature check would interpret this as an invalid block, but a partial check would not.
\end{quote}
Furthermore, we note that a generalization of this attack will work
\emph{regardless of the random coefficient size} if subgroup validation
is not done: if deviations $D_1$ and $D_2$ are chosen as element of an
order-$p$ subgroup, and signatures submitted are
$S_1=(C_1+D_1)$ and $S_2=(C_2+D_2)$, then $r_1 D_1=-r_2 D_2$ with chance
$1/p$, in which case the weighted deviations will cancel themselves out,
and verification will pass.

Details and further analysis appear in a recent post that we
contributed to~\cite{BLSfv}.

\subsection{Implementations}\label{ssec:blsimpls}

The main implementation of BLS signatures is the blst project\footnote{\url{https://github.com/supranational/blst}} (``blast'') from Supranational.
blst's BLS logic is written in C, with some x86\_64 and ARMv8 assembly for core arithmetic operations, and the project provides bindings for Go, and Rust, as well as partial support for Python, Java, and Node.js.  

blst is described as ``focused on performance and security'', and indeed shows good attention to security, with safety checks and a number of comments in the code related to security mitigations and design choices.
For example, the blst team attempts to document how to use its APIs and the users' responsibilities, as in the following:
\begin{quote}
    The essential point to note is that it's the caller's responsibility to ensure that public keys are group-checked with \code{blst\_p1\_affine\_in\_g1}. This is because it's a relatively expensive operation and it's naturally assumed that the application would cache the check's outcome. Signatures are group-checked internally.
\end{quote}
Different clients have a different approach, but most validate the key upon deserialization of the bytes object and creation of a public key object (see for example the discussion in \url{https://github.com/ConsenSys/teku/issues/4025}).
Worth noting too, blst warns users of confusing aspects of the API, noting for example that
\begin{quote}
unlike what your intuition might suggest, \code{blst\_sign\_*} doesn't sign a message, but rather a point on the corresponding elliptic curve
\end{quote}
and that
\begin{quote}
    Another counter-intuitive aspect is the apparent g1 vs. g2 naming mismatch, in the sense that \code{blst\_sign\_pk\_in\_g1} accepts output from \code{blst\_hash\_to\_g2}, and \code{blst\_sign\_pk\_in\_g2} accepts output from \code{blst\_hash\_to\_g1}. This is because, as you should recall, public keys and signatures come from complementary groups.
\end{quote}

blst was audited by NCC, covering all the code from the assembly arithmetic up to the Go and Rust bindings~\cite{NCCblst}, reporting mostly low-severity issues.
Third-party bindings exist for Java
(\href{https://github.com/ConsenSys/jblst}{ConsenSys/jblst} used in
Teku), Nim
(\href{https://github.com/status-im/nim-blscurve}{status-im/nim-blscurve},
used in Nimbus), and TypeScript
(\href{https://github.com/ChainSafe/blst-ts}{ChainSafe/blst-ts}).

\section{Slashing}\label{sec:slashing}

The \emph{slashing} mechanism aims to disincentivize ``bad'' behavior from the validators by applying penalties on the validator revenue and eventually excluding it from the network.
At the same time, slashing incentivizes reports of such behavior by offering a ``whistleblower reward'' to the validator that submits a proof of a validator's misbehavior.

Slashing is part of the Ethereum reward and penalty mechanism, but should not be confused with the penalty mechanism that punishes a node for being offline or for a miscast vote, which are unlikely to be malicious activities.
Instead, slashing punishes behavior that violates the protocol definition and that could potentially jeopardize the consensus' security.
Specifically, slashing punishes validators that
\begin{itemize}
    \item As proposers, propose different beacon blocks for the same slot
    \item As attesters, sign conflicting attestations
\end{itemize}
The high-level workflow is then the following: 
\begin{enumerate}
\item \emph{slasher} services monitor proposed blocks and attestations for invalid ones
\item When a slasher detects a slashable event among proposed blocks and
attestations, it communicates to a validator (other than the one guilty)
a \code{ProposerSlashing} or \code{AttesterSlashing} object
\item The validator submits the slashing into a block
\item Other validators verify the slashing proposal correctness, and if
the block is validated then the validator that proposed the slashing
receives a \emph{whistleblower reward} 
\end{enumerate}
A list of slashing events is available at \url{https://beaconscan.com/slots-slashed}.

The logic of Ethereum's slashing actions is defined\footnote{\url{https://github.com/ethereum/eth2.0-specs/blob/dev/specs/phase0/beacon-chain.md\#slash_validator}}, in
\code{slash\_validator()} and \code{process\_slashings()},
while processing of slashing objects is defined in
\code{process\_proposer\_slashing()} and
\code{process\_attester\_slashing()}.

\subsection{Protection}

Slashing happens when a slasher detect a behavior similar to a
potentially malicious one, but in practice such behavior may occur
accidentally rather than maliciously.
For example, validators may attempt to minimize downtime (and associated
penalties) by running multiple instances behind a load balancer, however
a faulty setup can lead to two instances proposing conflicting blocks.
The first slashing event allegedly happened for such a
reason\footnote{\url{https://beaconcha.in/validator/20075}}.
Accidental slashing can be in part due to a missing or compromised
attestation history, for example, when migrating a validator's database.
The Ethereum specifications include some recommendations, such
as\footnote{\url{https://github.com/ethereum/eth2.0-specs/pull/2107/files}}:

\begin{quote}
before a validator client signs a message it should validate the data,
check it against a local slashing database (do not sign a slashable
attestation or block) and update its internal slashing database with the
newly signed object.
\end{quote}
EIP-3076~\cite{eip3076} (``Slashing Protection Interchange Format'') was created
to prevent such accidental slashing, by defining a standard JSON-based record format to migrate
validator histories across instances' of a beacon client.
All the beacon clients reviewed support EIP-3076: Lighthouse and Nimbus store data in an SQLite database, Teku in a YAML file, and Prysm in the validator's BoltDB instance.
Validators then attempt to prevent accidental slashing by identifying
contradictions between new events and previously recorded events (proposals,
attestations) to prevent accidental slashable events.
Note that different clients may implement slightly different policies\footnote{\url{https://ethereum-magicians.org/t/eip-3076-validator-client-interchange-format-slashing-protection/4883/2}}.

Such protections work when multiple client instances use the
same data directory (datadir) and database, but are ineffective if
instances run independently with the same key,
for example, failover instances on another infrastructure.

Given EIP-3076-based protection, as users get more experienced, and risks are
better documented, we can expect accidental slashings to vanish on the long
run.

\subsection{Security Model}

The security goals of the slashing mechanism are to disincentivize
adversarial abuse, and to prevent circumvention of the detection
mechanism:

\begin{enumerate}
    \item A dishonest validator should not be able to misbehave in such
    a way that they can't be detected (and then slashed), or that
    another (innocent) validator is slashed instead of them. This would
    happen if all slashers were ineffective at a given epoch, for
    example.

    \item A malicious validator should not be capable of
    enticing another validator into unknowingly committing a slashable
    offense, so that the malicious validator receives the whistleblower
    reward.  This may happen if circumstances are such that
    a validator switches to a failover instance with no slashing
    protection.

\end{enumerate}
Examples of attacks on slashing have been
described\footnote{\url{https://ethresear.ch/t/global-slashing-attack-on-eth2/6703}}.
However, under the assumptions described, an attacker can do more damage than just collecting whistleblower rewards or blackmailing the validator.

Additionally, potential issues can come from
\begin{itemize}
    \item Errors, such as miscalculations of rewards, when receiving multiple slashing attestations, and including multiple in a block (note that \code{MAX\_ATTESTER\_SLASHINGS=2} and \code{MAX\_PROPOSER\_SLASHINGS=16})

    \item Leverage of invalid slashing attestations for malicious purposes; note that a validator proposing an invalid slashing attestation is not penalized, let alone slashed.
\end{itemize}
Failures of slashing can come from: 
\begin{itemize}
    \item Flaws in the slashing logic, as specified, 
    \item Implementation errors of said logic, or 
    \item Via the creation of a state under which slashing is ineffective. 
\end{itemize}
Our review focused on the second aspect, the implementation, which consists mainly in:
\begin{itemize}
    \item Detection of slashing conditions, by slasher services.
    (See~\cite{slashdec} for detection methods,
    and~\cite{weaksub,weaksub2} for fundamental analysis and the concept
    of weak subjectivity.)

    \item Processing of slashing attestations, by validators.
\end{itemize}

\subsection{Implementations Review}\label{slashing:impl}

In all clients' validator implementations, we checked the correct
implementation of the specification's \code{is\_slashable\_validator()},
\code{is\_slashable\_attestation\_data}, \code{slash\_validator()},
\code{process\_slashings()}, \code{process\_attester\_slashing()},
\code{process\_proposer\_slashing()}, as well as use of the correctness
of the constants' values.

We found implementations to be consistent with the specifications, and
only noted that Nimbus' implementation can bypass signatures verification
in  \code{check\_proposer\_slashing()} (which implements the logic of the specs'
\code{process\_proposer\_slashing()}) when the \code{skipBlsValidation}
is set. 
This flag seems to only be used for test routines, but otherwise it
would allow the slashing of innocent validators (from the perspective of
a Nimbus validator).

We did not find flaws in Teku's slashing implementation, but it's the
implementation we found the hardest to parse, and the one we are the
least confident in.

\medskip

\noindent
{\large \Warning}\textbf{Recommendation:}
Review the way slasher services detect slashing conditions, and how it
could be abused. (We did not thoroughly review this part.)


\section{P2P Networking}\label{sec:noise}

Ethereum uses~\cite{eth2net} the libp2p~\cite{libp2p} multi-transport
stack for secure transport between peers. 
Specifically, Ethereum uses the libp2p-noise~\cite{libp2p-noise}
Noise-based~\cite{noise} protocol, which superseded the SECIO protocol\footnote{\url{https://blog.ipfs.io/2020-08-07-deprecating-secio/}}.

libp2p-noise uses the Noise~XX pattern over the curve Curve25519, with a
few twists.
libp2p-noise bootstraps from long-term \emph{identity keys}, as opposed
to Noise static keys, and updates Noise static keys regularly, signing
new keys with the identity key, and including this signature in a
handshake payload.

The Ethereum networking specification~\cite{eth2net} defines the following network security goals:
\begin{itemize}
    \item Peer authentication
    \item Confidentiality
    \item Integrity
    \item Non-repudiation
    \item Non-replayability
    \item Perfect forward secrecy
\end{itemize}
Let's review to what extent these are satisfied.

\subsection{libp2p-noise Implementations}

At the time of writing, libp2p has 7 native
implementations~\cite{libp2p-impl} including implementations in Go,
Rust, Nim, TypeScript, and one in Java.
There are three ways to implement a Noise pattern:
\begin{itemize}
\item Implement a Noise version (called ``pattern'') from scratch
\item Use a reference implementation in the corresponding programming language
\item Use the automatically generated template for a target pattern 
\end{itemize} 
Prysm, Lighthouse and Teku use libp2p-noise versions based on the
corresponding reference implementations of Noise:
flynn/noise\footnote{\url{https://github.com/flynn/noise}},
mcginty/snow\footnote{\url{https://github.com/mcginty/snow}}, and
rweather/noise-java\footnote{\url{https://github.com/rweather/noise-java}},
respectively.
Nimbus' libp2p-noise reimplements Noise~XX pattern from
scratch\footnote{\url{https://github.com/status-im/nim-libp2p/blob/master/libp2p/protocols/secure/noise.nim}},
which is arguably riskier than using an established implementation.
Lodestar uses a Noise~XX implementation generated from Noise~Explorer.

We reviewed the implementations' security and correctness with respect
to the Noise~XX specification, using as a baseline the flynn/noise Go
package, recently
audited\footnote{\url{https://cure53.de/pentest-report_turbotunnel.pdf}})
and the Noise specification document.  
We discovered a nonce overflow issue in the js-libp2p-noise
library\footnote{\url{https://github.com/NodeFactoryIo/js-libp2p-noise}}
(used in the Lodestar client) , which we traced back to a bug in
Noise~Explorer\footnote{\url{https://noiseexplorer.com/patterns/XX/}},
which js-libp2p-noise is using.
The problem is the following:
the Noise specification\footnote{\url{https://noiseprotocol.org/noise.html\#the-cipherstate-object}} says in \S5.1 that 
\begin{quote}
n: An 8-byte (64-bit) unsigned integer nonce. The maximum n value ($2^{64}-1$) is reserved for other use. If incrementing n results in $2^{64}-1$,
 then any further EncryptWithAd() or DecryptWithAd() calls will signal an error to the caller.
\end{quote}
However, the Noise~Explorer templates use the following code:
{
\footnotesize
\begin{minted}{javascript}
    type cipherstate struct {
        k [32]byte
        n uint32
    }
    (...)
    func encryptWithAd(cs *cipherstate, ad []byte, plaintext []byte) (*cipherstate, []byte) {
        e := encrypt(cs.k, cs.n, ad, plaintext)
        cs = setNonce(cs, incrementNonce(cs.n))
        return cs, e
}
\end{minted}
}

The lack of integer overflow check combined with the shorter nonce can
cause the session key and nonce to be reused after multiple messages. 
Despite the fact that js-libp2p-noise uses the ``number'' type instead of real \code{uint32} type, the implementation is still vulnerable
since nonces are converted to bytes\footnote{\url{https://github.com/NodeFactoryIo/js-libp2p-noise/blob/master/src/handshakes/abstract-handshake.ts\#L50}} using the fixed number of bytes.
Noise~Explorer acknowledge the issue and fixed it.

\subsection{libp2p-noise Protocol Security}

\subsubsection{Static Keys Signatures Replay}\label{ssec:noisesign}

Libp2p-noise extends the Noise~XX pattern by introducing \emph{identity
keys}, or long-term keys, and using Noise static keys 
between ephemeral and long-term keys: implementations may generate a new
static keypair for each session or a single static keypair may be
generated when libp2p-noise is initialized and then used for all sessions.

To authenticate the static key used in the Noise~XX handshake,
libp2p-noise includes in the handshake protocol
$\mathsf{NoiseHandshakePayload}=\mathsf{(id\_key, id\_sig, data)}$
message containing a signature of the static public key computed with
the identity private key: in Noise terms $\mathsf{id\_sig} =
\mathsf{Sig}(\mathsf{id\_key}, \mathsf{data})$, where $\mathsf{id\_key}$ is the sender's identity
private key and $\mathsf{data}$ is the ``noise-libp2p-static-key:'' string
followed by sender's static public key $\mathsf{s\_pub}$
\footnote{\url{https://github.com/libp2p/specs/tree/master/noise\#the-libp2p-handshake-payload}}.

The signature is computed over the sender's static public key without any unpredictable challenge from the corresponding peer. 
So, the used static key authentication mechanism violates the basic authenticated key agreement protocol design principle:
``Ephemeral leakage should not allow for long-term impersonation''~\cite{ake}.
In this case, the sender proves knowledge of the signature over the static public key, but not access to the identity private key within the current session.
For example, if an attacker finds a triple $(\mathsf{s\_pub},
\mathsf{s\_priv}, \mathsf{Sig(id\_key, s\_pub}))$ for the identity key
$\mathsf{id\_key}$ of the target user then they will be able to
impersonate this user without any limitation in the future.  As a
result, if an attacker is able to sign a static public key once then
they will be able to impersonate the identity key owner forever.  This
may occur if the attacker has temporary access to a signing module, or
if the static key and the signature are leaked to the attacker.

\medskip

\noindent
{\large \Warning}\textbf{Recommendation:}
Consider implementing mitigation such as using a peer's ephemeral public key
($\mathsf{re}$) as an unpredictable challenge in signing $\mathsf{data}
=
\mathit{"noise\textnormal{-}libp2p\textnormal{-}static\textnormal{-}key:"}\:||\:\mathsf{re}\:||\:s$.

\subsubsection{No DoS Countermeasures} 

Libp2p was designed to support multiple transport protocols (TCP, UDP, QUIC, etc.).
All datagram-based secure transport protocols (e.g., DTLS, IPsec, WireGuard) provide protection 
against denial-of-service (DoS) attacks.
For instance, WireGuard's Noise~IK-based protocol uses cookies to authenticate a session initiator and has a second message that is smaller than the first message to prevent amplification attacks. 

No such defense is implemented in libp2p-noise, thus an attacker could flood the responder with session initiation messages and force them to compute exponentiations.
The initiator could also spoof its origin address and exploit the data amplification provided by Noise~XX,
 whose first is 32-byte and the second is 192-byte, or a potential 6$\times$ amplification factor (without early data).

 We believe that at present time this risk is mitigated by TCP mechanisms.


\subsubsection{Early Data Insecurity} 

The libp2p-noise specification states the following related to early data (payloads):
``These payloads MUST be inserted into the first message of the handshake pattern that guarantees secrecy.
In practice, this means that the initiator must not send a payload in their first message.
Instead, the initiator will send its payload in message 3 (closing message),
whereas the responder will send theirs in message 2 (their only message)''
 
According to Noise~XX security properties~\cite{noise}, the second message with payload provides forward secrecy,
however, the sender has not authenticated the responder, so this payload might be sent to any party,
including an active attacker. So an active attacker can just establish a connection with the responder host and get the early data.

\subsubsection{Identity Hiding} 

The Ethereum specification doesn't mention identity hiding as a
security goal, but we observed that it inherits identity hiding from
Noise~XX.
many Noise patterns by design including Noise~XX.
The payload security and identity-hiding properties of the original
Noise~XX handshake pattern are as follows:
\begin{itemize}
    \item The responder's static public key is encrypted with forward secrecy but can be probed by an anonymous initiator
    \item The responder's handshake payload is encrypted with forward secrecy, depending on an ephemeral key, but the payload might be sent to any party
    \item The initiator's static public key is encrypted with forward secrecy to an authenticated party
    \item The initiator's third handshake message payload is encrypted with forward secrecy for an authenticated party
\end{itemize}
The static public keys and handshake payloads in libp2p-noise thus have
similar security properties.
Note however that libp2p-noise transmits identity keys in handshake
payloads, and uses them instead of Noise temporal static keys, while
statics keys are updated for each session (and are thus not long-term peer
identifiers).
The identity hiding of Noise~XX with respect to static keys thus
applies as well to identity keys in libp2p-noise.

\subsubsection{No Non-Repudiation}

Although non-repudiation is stated as a security goal, 
libp2p-noise does not provide non-repudiation for transport messages.
This would require digital signatures with the sender's (identity,
static, or ephemeral) key, however only the second and third handshake
messages are signed.

\subsection{discv5 Handshake}\label{ssec:ndp}

Ethereum consensus nodes~\cite{eth2net}  discover each other using
the Node Discovery Protocol Version 5.1 (discv5)~\cite{discv5}.
discv5 is a UDP protocol that works with self-certified, flexible Ethereum node
records (ENRs) and topic-based advertisement, both of which are
requirements in this context.
At the time of writing, the discv5 specification is still a work in progress.

discv5 aims to provide:
\begin{itemize}
  \item Network traffic encryption to protect against passive observers
  \item Authentication, insofar as peers are known and trusted on a TOFU basis
  \item Network traffic obfuscation to prevent traffic mangling, naive blocking of the protocol messages using hard-coded packet signatures, and trivial sniffing
\end{itemize}
Most of the design requirements and security goals of discv5 are devoted
to the protocol application logic (e.g., Kademlia redirection, replay of 
NODES or PONG response packets, traffic amplification, Sybil/eclipse
attacks), and don't address secure transport mechanisms. 
An implicit  assumption is that all communications should be encrypted
and authenticated, protecting topic searches and record lookups against
passive observers.
The discv5 design document states the following:
\begin{itemize}
  \item The handshake protocol protects against passive observers but is not forward-secure and active protocol participants can access node information by simply asking for it
  \item Since the handshake performs cryptographic operations (ECDH, signature verification for different algorithms) performance of the handshake is a big concern
  \item discv5 handshake reduces the risk of computational DoS because it costs as much to create as it costs to verify and cannot be replayed.
\end{itemize}

\subsubsection{Protocol}

This is a simplified description of the discv5 handshake from a
cryptography perspective, based on its
specification\footnote{\url{https://github.com/ethereum/devp2p/blob/master/discv5/discv5-theory.md}}.
The handshake protocol involves the sending of the following discv5
messages, whose headers include a fixed-length field \code{static
header} defined as \code{protocol-id || version || flag || nonce ||
authdata-size}, where \code{nonce} is a random 96-bit value, used for
AES-GCM authenticated encryption.

In the following, node A (initiator) wishes to communicate with node B (responder). Node A knows node B's
identity (that is, its identity key). 
In our description, B does not know A and never communicated with A.

\begin{enumerate}
  
  \item A sends its identity key and a nonce to B, as a FINDNODE
  message.
  
  \item B initiates the actual handshake by sending as a challenge a
  WHOAREYOU packet comprising:
  \begin{itemize}
    \item A random 128-bit \code{id-nonce} field.
    \item A sequence number field \code{enr-seq} set to zero.
  \end{itemize}
  
  \item A then:
  \begin{itemize}
    \item Generates an ephemeral key pair $(\mathtt{ephemeral\text{-}key},
    \mathtt{ephemeral\text{-}pubkey})$ (using the specs' notations).
   
   \item Uses B's identity key \code{dest-pubkey} to derive session keys from
    $\ECDH(\mathtt{ephemeral-key}, \mathtt{dest-pubkey})$ (using HKDF).

    \item Computes the \code{signature} field using its identity private
    key (and the elliptic-curve signature scheme used by the instance),
    signing a message comprising the challenge data (as sent by B),
    \code{ephemeral-pubkey}, and B's identifier.

    \item Sends \code{ephemeral-pubkey}, \code{signature}, and a payload
    encrypted with the derived keys.
  \end{itemize}

 \item B verifies \code{signature}, derives keys and decrypts the payload, and responds with a
 NODES message (encrypted using the derived \code{recipient-key}).

 \item A verifies that it can decrypt B's message, and then considers
 B's identity verified and the session keys valid.
 
\end{enumerate}

\subsubsection{Security}

The discv5 specification claims that protocol messages are secure
against ``passive observers''.
However, the security goals and design rationale are unclear, in
particular regarding:
\begin{itemize}
  \item The authentication asymmetry: a responder uses static
  Diffie-Hellman, while the initiator uses signatures.
  
  \item The use of static Diffie-Hellman (and no ephemeral-ephemeral
  combination).  An authenticated key agreement doesn't necessarily
  require static Diffie-Hellman, a signature and ephemeral
  Diffie-Hellman can be used instead.

  \item If the protocol considers a passive attacker only then the
  handshake protocol may not use authentication at all, since
  Diffie-Hellman key agreement is secure against a passive attacker but
  insecure against an active attacker.
\end{itemize}
It follows that forward secrecy holds for sender compromise only: if the responder's static private key is compromised, the past messages can be decrypted.
Moreover, a passive observer can decrypt all future messages sent by a sender to the responder on the fly.

We believe that the protocol's security can be improved without any
changes in the current flow or RTT number and slightly affecting
performance.  As an illustration, we show how the Noise~KK pattern can be
used as a basis.

First, note that ``Ordinary message packet''\footnote{\url{https://github.com/ethereum/devp2p/blob/master/discv5/discv5-wire.md\#ordinary-message-packet-flag--0}}
contains A's identifier \code{id-nonce} corresponding to the static public key. So, on the step 3, A and B knows identities (static public keys) of each other.
So, the new protocol can be defined as follows:
\begin{enumerate}

\item A sends its identity key and a nonce to B, as a FINDNODE message

\item B initiates the actual handshake by sending as a challenge a
  WHOAREYOU packet comprising:
  \begin{itemize}
    \item A random 128-bit \code{id-nonce} field.
    \item A sequence number field \code{enr-seq} set to zero.
  \end{itemize}

\item A then:
  \begin{itemize}
    
    \item Generates an ephemeral key pair $(\mathtt{source\text{-}ephemeral\text{-}key},
    \mathtt{source\text{-}ephemeral\text{-}pubkey})$.
    \item Performs $\ECDH(\mathtt{source\text{-}ephemeral\text{-}key},
    \mathtt{dest\text{-}pubkey})$, $\ECDH(\mathtt{source\text{-}key},
    \mathtt{dest\text{-}pubkey})$.
    \item Mixes the outputs of DH using HKDF and derives session keys.
    \item Computes \code{signature} using its identity private key, signing a
    message comprising the challenge data (as sent by B),
    \code{source-ephemeral-pubkey}, and B's identifier.

    \item Sends \code{source-ephemeral-pubkey}, \code{signature}, and a payload
    encrypted with the derived keys.
  \end{itemize}

\item B:
  \begin{itemize}
    \item Verifies \code{signature}, performs
    $\ECDH({\mathtt{dest\text{-}key}},
    \mathtt{source\text{-}ephemeral\text{-}pubkey})$, \\
    $\ECDH(\mathtt{dest\text{-}key}, \mathtt{source\text{-}pubkey})$,
    derives keys and decrypts the payload.
    \item Generates an ephemeral key pair
    $(\mathtt{dest\text{-}ephemeral\text{-}key},
    \mathtt{dest\text{-}ephemeral\text{-}pubkey})$.
    \item Performs $\ECDH(\mathtt{dest\text{-}ephemeral\text{-}key},
    \mathtt{source\text{-}ephemeral\text{-}pubkey})$, \\
    $\ECDH(\mathtt{dest\text{-}key},
    \mathtt{source\text{-}ephemeral\text{-}pubkey})$.
    \item Mixes the outputs of DH using HKDF and derive session keys.
    \item Responds with \code{dest-ephemeral-pubkey} and a NODES message (encrypted using the derived key).
  \end{itemize}

\item A:
  \begin{itemize}
    \item Receives B's ephemeral public key, performs
    $\ECDH(\mathtt{\mathtt{source\text{-}ephemeral\text{-}key},
    dest\text{-}ephemeral\text{-}pubkey})$, \\
    $\ECDH(\mathtt{source\text{-}ephemeral\text{-}key},
    \mathtt{dest\text{-}pubkey})$.
    \item Verifes that it can decrypt B's message, and then considers
    B's identity verifed and the session keys valid.
  \end{itemize}
\end{enumerate}

This pattern seems to provide the highest security
guarantees (number 2 and 5) in Noise terms for all next messages (e.g.,
FINDNODE, PING, PONG, etc.), but a more thorough analysis is needed if
discv5 will consider using it.

Another concern is that ``sig-size'' (signature size), and ``eph-key-size'' (ephemeral key size) fields of a handshake message's ``authdata-head'' are not authenticated:
the values are not used either in key agreement or signing. 
At the time of writing, ``sig-size'' and ``eph-key-size'' are constants in the v4 scheme.
The traditional approach is to compute the handshake transcript hash, by hashing the concatenation of all messages sent and received during a handshake.
If both sides do not compute the same transcript hash, the connection
must be aborted.


\subsection{Gossipsub Peer Scoring}\label{gossip}

Gossipsub is an extensible publish/subscribe protocol over libp2p. 
The v1.0\footnote{\url{https://github.com/libp2p/specs/blob/master/pubsub/gossipsub/gossipsub-v1.0.md}} implements a publish/subscribe model in peer-to-peer networks.
The v1.1\footnote{\url{https://github.com/libp2p/specs/blob/master/pubsub/gossipsub/gossipsub-v1.1.md}} is a set of security extensions addressing protocol attacks.
In June 2020, Least Authority completed an audit\footnote{\url{https://leastauthority.com/blog/audit-of-gossipsub-v1-1-protocol-design-implementation-for-protocol-labs/}}
of the Gossipsub v1.1 design and its implementation built on the libp2p library.
The report identified the peer scoring mechanism as high risk for the following main reasons:
\begin{itemize}
  \item A node can leak scoring information
  \item The peer scoring mechanism can adjust the network and increase its centralization 
\end{itemize}
The risk is that the peer scoring mechanism be abused to prevent
detection of malicious nodes, or flag honest nodes as malicious or
unreliable.
The current mechanism is similar to a linear regression model:
weights (coefficients) are estimated by an algorithm
based on the past behavior of the system.
However, the weights of the score function are theoretically
evaluated, the violation thresholds are fixed heuristically.
For instance, the Lighthouse team writes\footnote{\url{https://hackmd.io/FxenPiVmT5WR3c7eScR-gA}}: 
\begin{quote}
These are initial values based on theoretically expected numbers and are likely to change during further simulations
(...)
We theoretically calculated expected scoring parameters based on
reasonable network variables (such as expected delay between packets,
expected duplicates, rate of messages, and topic sizes).
\end{quote}
It is hard to determine and empirically evaluate to what extent such
weights are ``good enough'' to prevent abuse and to adequately model the
notion of good behavior in order to maintain a reliable system.
Theoretical estimates risk not representing reality
accurately enough, while purely empirical ones risk 
``overfitting'' by expecting future behavior to reflect the past.

Evaluation and simulations validated the soundness of the proposed
model, but may not be sufficient in an adversarial scenario.
We shared our concerns with the authors of the original Gossipsub
research~\cite{GOSSIPSUB}.

\section{API Implementations}\label{sec:api}

Beacon clients expose the Beacon Node API~\cite{eth2api} for querying the
beacon node, as API used by validators to determine their assigned duties, submit block
proposals, etc.
Each beacon client has its own specificities regarding the API:
\begin{itemize}
    \item Nimbus uses JSON-RPC
    2.0\footnote{\url{https://github.com/status-im/nimbus-eth2/blob/stable/docs/the\_nimbus\_book/src/api.md}}
    and HTTP-JSON interfaces, and listed some security
    recommendations\footnote{\url{https://github.com/status-im/nimbus-eth2/issues/1665}}.
    \item Lighthouse implements the standard and non-standard RESTful
    APIs\footnote{\url{https://lighthouse-book.sigmaprime.io/api-lighthouse.html}}
    for the beacon node and the validator
    client\footnote{\url{https://lighthouse-book.sigmaprime.io/api-vc.html}}.
    \item Prysm implements the API (used only by Prysm) using
    gRPC\footnote{\url{https://docs.prylabs.network/docs/how-prysm-works/ethereum-2-public-api/}}
    and also exposes HTTP-JSON version of the API.
    \item The both API services of Prysm may be secured by TLS with
    default cipher suites.
    \item Teku implements additional API endpoints (e.g., log\_level,
    peer\_scores, etc.).
\end{itemize}


\subsection{Requests Validation}

We evaluated the API implementation based on industry-standard best practices, notably
relying on the OWASP lists~\cite{owaspapi,owaspasvs} of security
controls\footnote{\url{https://github.com/OWASP/ASVS/blob/master/4.0/en/0x21-V13-API.md}}. 
Applicable controls from this reference include for example signaling
errors with appropriate response statuses, checking Content-Type
correctness, and so on.
The Ethereum specifications\cite{eth2spec,eth2api} says little about
the security requirements of the API, but mentions these aspects: ``All
requests by default send and receive JSON, and as such should have
either or both of the "Content-Type: application/json" and "Accept:
application/json" headers.'' 
However, not all clients do this: 
\begin{itemize}
  \item Teku handles requests with arbitrary Content-Type and
  Accept headers ignoring them and using application/json.
  \item Lighthouse rejects requests with wrong Content-Type header but
  accepts arbitrary Accept headers and also uses application/json.
  \item Prysm ignores those headers.
\end{itemize}
Another requirement is that ``JSON schema validation is in place and verified before accepting input'':
\begin{itemize}
  \item Teku performs JSON validation and responds with ``Unrecognized
  field'' error message if unknown fields are found.
  \item Lighthouse and Prysm accept requests with unknown fields and
  process input ignoring them, at the same time they do respond with a
  deserialize error message if a field value has a wrong type.
\end{itemize}


\subsection{Exposure}

The beacon node API specification~\cite{eth2spec,eth2api} does not
specify authentication requirements, and claims that the API is public:
``The API is a REST interface, accessed via HTTP, designed for use as a public communications protocol''\footnote{\url{https://github.com/ethereum/eth2.0-APIs\#outline}}.
Exposing the API publicly leaves the service potentially vulnerable to
external attacks and abuse, and is in general not necessary.
Some clients thus recommend to only serve the API locally, or to
authorized hosts, for example:
\begin{itemize}
\item Teku: ``Only trusted parties should access the REST API. Do not directly
expose these APIs publicly on production nodes.''\footnote{\url{https://docs.teku.consensys.net/en/latest/Reference/Rest_API/Rest/}}:
\item Lighthouse: ``the API should only be exposed to
localhost or a restricted set of IPs on an internal
network''\footnote{\url{https://github.com/sigp/lighthouse/issues/2468\#issuecomment-882936767}},
and ``Do not expose the beacon node API to the public internet or you
will open your node to denial-of-service (DoS)
attacks.''\footnote{\url{https://lighthouse-book.sigmaprime.io/api-bn.html\#security}}.
\end{itemize}
We nonetheless identified a few hosts exposing the API publicly, by
scanning IPv4 addresses for beacon API endpoints on the ports known to
be used for this API (e.g., 3500, 5051, 5052). We found 20 Lighthouse
instances, 0 Nimbus, 5 Prysm, and 5 Teku. 
See~\S\ref{sec:fngp} for methodology details.


\subsection{Authentication}\label{ssec:apiauth}

Prysm protects gRPC connections using
TLS\footnote{\url{https://docs.prylabs.network/docs/prysm-usage/secure-grpc/}},
allowing a validator to authenticate a beacon node (but not the other
way).
Otherwise, clients don't provide authentication mechanisms to restrict
access to the beacon node API.
At best, they inform users via a warning in the documentation that the API must not be exposed to untrusted parties.
Note that in some contexts, server-side request forgery (SSRF) could be
exploited to access the beacon node API, if an authorized service is
vulnerable to SSRF.

Lighthouse uses logging to deliver authentication tokens (described
below) for Web UI, and Prysm has a feature request~\footnote{\url{https://github.com/prysmaticlabs/prysm/issues/9188}},
suggesting the same approach (``the validator client can generate a
random auth token and log it to stdout + write it to a file to persist
it''). 
This approach is insecure by design and is considered known as unsafe
practice in web application security.

\subsection{Lighthouse Validator Client API}\label{vcapi}

Lighthouse implements a custom API, called Validator Client
API~\cite{VC}.
Since the validator client (VC) can be used to access validator keys,
the GUI accessing it must be authenticated. 
The security protocol between a VC and a
browser-based GUI is described in a GitHub
issue\footnote{\url{https://github.com/sigp/lighthouse/issues/1269\#issuecomment-649879855}} and in
the Lighthouse book~\cite{VC}.
In this protocol, the VC has a key pair and signs its
responses (ECDSA-secp256k1), while the public key is passed by the GUI in the request.
Note that these two references differ slightly: the former requires a
signature over HTTP response headers and body, but the latter for its body only.

A problem with this protocol is that replays of signed responses are
possible, because the signed data does not include an unpredictable
challenge or a timestamp. Adding a
``Date'' header in the HTTP headers would partially mitigate this.

We also reported some bugs in the implementation of the protocol:
\begin{itemize}
   \item The implementation doesn't require an API token for POST and
   PATCH
   requests\footnote{\url{https://github.com/sigp/lighthouse/issues/2512}}.
   \item A signature is computed over HTTP response body without
   headers\footnote{\url{https://github.com/sigp/lighthouse/issues/2511}}.
   \item API keys\footnote{\url{https://github.com/sigp/lighthouse/issues/2437}} and tokens\footnote{\url{https://github.com/sigp/lighthouse/issues/2438}} are stored locally with insufficient permissions restrictions (644).
 \end{itemize}

\subsection{API Denial of Service}\label{apidos}


The Ethereum networking specification~\cite{eth2net} comments on
potential DoS vectors and corresponding protection.
For instance, it contains a \code{BeaconBlocksByRange} request/message
potentially vulnerable to DoS. 
All clients have already implemented or are implementing\footnote{\url{https://github.com/status-im/nimbus-eth2/issues/1359}} 
rate limiting for P2P network and RPC mechanisms.


However, we found other endpoints vulnerable to DoS and not protected in
clients' implementations:
For example, /eth/v1/validator/duties/attester/\{epoch\} requests the beacon node to provide a set of attestation duties, which should be performed by validators,
for a particular epoch. Its request body contains an array of the validator indices for which to obtain the duties. 
So epoch and validator indices parameters directly affect performance and liveness.

We used two tests to assess response time. Tests were performed on clients with default settings.

In the first test, we send requests where the payload is a big array of the validator indices for which to obtain the duties (e.g., an array with $n$ indices $[1,2, ..., n]$, where n is 300, 600, 1000)
and the epoch is the current epoch. 
We observed that some client nodes responded with delay (e.g., for 830 indices payload the delay was 100 seconds), others stopped responding to all requests for a long time. 

{
\footnotesize
\begin{verbatim}

time curl -i -s -k -X 'POST' \
  -H 'accept: application/json'
  -H 'Content-Type: application/json'
  -d '[$Payload]' \
'http://$BeaconNodeAPIHost:$Port/eth/v1/validator/duties/attester/$CurrentEpoch'

\end{verbatim}
}

The second test, where the payload contains one index, but the epoch is a historic epoch. 
For instance, we observed requests completed in more than 40 seconds if the distance between epochs is 5 (e.g., the current epoch is 64555, the old epoch is 64550).
If the distance between epochs is 155 (e.g., the current epoch is 64555, the old epoch is 64400) the response time was about 3 minutes.
The problem here is that it is necessary to load historic data from a database.
{
\footnotesize
\begin{verbatim}

time curl -i -s -k -X 'POST' \
  -H 'accept: application/json'
  -H 'Content-Type: application/json'
  -d '[1]' \
'http://$BeaconNodeAPIHost:$Port/eth/v1/validator/duties/attester/$OldEpoch'

\end{verbatim}
}

We reported this issue to Lighthouse\footnote{\url{https://github.com/sigp/lighthouse/issues/2468}}, Nimbus\footnote{\url{https://github.com/status-im/nimbus-eth2/issues/2734}},
and Prysm\footnote{\url{https://github.com/prysmaticlabs/prysm/issues/9247}}.
There are beliefs that such attacks can not be prevented by input validation in the considered application domain.
Because of that, the issue is a good example of why authentication is needed as a mitigation mechanism
and what an attacker can perform if the beacon node API is accessible on the internet. 

\subsection{BLS Remote Signer HTTP API}\label{sec:remotesigner}

This simple API\footnote{\url{https://eips.ethereum.org/EIPS/eip-3030}} allows
a validator client to request signatures to a service storing private
keys, such as a key vault or a hardware security module (HSM).
In most deployments, such a signing service should explicitly
authenticate the requester, to prevent signing data from any party.

The API has been implemented by 
  Teku\footnote{\url{https://docs.teku.consensys.net/en/latest/Tutorials/Configure-External-Signer-TLS/}},
  Prysm\footnote{\url{https://docs.prylabs.network/docs/wallet/remote/}},
  and
  Lighthouse\footnote{\url{https://github.com/sigp/lighthouse/tree/stable/remote_signer}}.

Lighthouse does not yet implement security mechanisms, Teku implements
mutual authentication (mTLS), and Prysm's documentation says it also
authenticates both parties, however the implementation only
authenticates the services. 
We
reported\footnote{\url{https://github.com/prysmaticlabs/remote-signer/issues/14}}
the latter issue.

\section{Supply Chain Risk Analysis}\label{sec:deps}

Software vulnerabilities can be in the project's own code, or in any
component of its ``supply chain'', that is, third-party code that the
application depends on, down from the CPU microcode, hypervisor,
operating system, language runtime, external
APIs, but mainly from explicitly called packages, modules, libraries,
which we'll just call \emph{dependencies}.
In our context, as far as we know, all dependencies are
open-source\footnote{Closed-source dependencies bear different types of
risks, which have been discussed at length in multiple articles and
posts.}.

External dependencies increase the security risk, because of:
\begin{itemize}
    \item The sheer amount of dependencies (and thus of code) in modern
    applications; the more code, the more bugs.
    \item The automatic download and update from remote hosts which, although
    authenticated, can break compatibility, introduce new pre-conditions
    or security limitations.
    \item The uneven maturity of dependencies' development lifecycle,
    and the varying reliability of their test suites.
    \item The common open-contribution model, where changes can be
    proposed by any stranger, with often little quality assurance or accountability
    \item The lack of guarantee or liability of any kind, as typically stated in open-source licenses
    \item Security audits of an application usually not covering its dependencies
    \item Application developers' testing framework focused on the application's code 
    \item Many abandoned, deprecated, unmaintained projects
    \item Version pinning to vulnerable and outdated components~\cite{DecMensCon018} 
\end{itemize}
The risk can then materialize as:
\begin{itemize}
    \item A greater density of bugs in dependencies than in the parent application
    \item A greater delay between a bug identification/reporting and patching in dependencies
    \item Active sabotage, such as via ``hypocrite commits''~\cite{HC}
    \item ``Dependency confusion'' attacks\footnote{\url{https://medium.com/@alex.birsan/dependency-confusion-4a5d60fec610}} 
    \item Software ``supply chain'' attacks injecting malicious code into a software package to compromise the dependent
    systems~\cite{VuPaMaPlaSa,OhmPS020}
    \item Typosquatting and ``combosquatting'' attacks through package manager ecosystems~\cite{TyCom,TaylorVDCR20}
\end{itemize}
To better understand this risk, we propose a number of indicators that
we considered in our evaluation of the beacon clients, but that are generally applicable to software projects.
We describe how to partially automate the collection of indicators, 
then discuss the values observed and the limitations of our approach,
finally, we offer concrete recommendations for beacon clients
developers. 

\subsection{Related Work}

The security assurance level of modern software is scary:
``80\% or more of most applications’ code comes from
dependencies'', reported GitHub's 2020 Octoverse
study~\cite{GITOCTO2020}.
Synopsys reported~\cite{SYNOP2021} that 85\% of
audited projects contained components that were outdated for over four
years or inactive for over four years. 
The 2020 Linux Foundation \& Harvard FOSS contributor survey revealed
that only 32\% of respondents used dependency analysis tools and that
security measures (signed commits, 2FA) are rarely enforced.


Several initiatives were created to raise awareness and reduce the risk associated to
dependencies, for example:
\begin{itemize}

\item Google created the Open Source Vulnerabilities (OSV) platform
\url{https://osv.dev}, which offers an API to query if a given version
of a component has known vulnerabilities.
Google also created the Open Source Insights tool
at \url{https://deps.dev}, which provides information about the
dependencies, including security advisories, and dependency graphs.

\item GitHub maintains the \url{https://github.com/advisories} database of
security advisories and offers extensive documentation about supply
chain
security\footnote{\url{https://docs.github.com/en/code-security/supply-chain-security}},
including dependency graph and automatic version update (with
Dependatbot).

\item The Linux Foundation created \url{http://sigstore.dev}, to provide free
certificates and tools to automate and verify signatures of software
components, to defend software supply chain attacks. 

\item OWASP provides the Dependency Check
platform\footnote{\url{https://owasp.org/www-project-dependency-check/}},
a tool `` tool that attempts to detect publicly disclosed
vulnerabilities contained within a project’s dependencies''

\end{itemize}
In addition to these, there are a lot of commercial and free open-source
software composition analysis tools (such as Synopsys's Black Duck),
which will attempt to inventory open-source dependencies and more
generally identify third-party code, in order to match it against
databases such as the National Vulnerability
Database\footnote{\url{https://nvd.nist.gov/}}.

\subsection{Characterizing Dependencies}\label{sec:deps:metrics}

In this section, we list the characteristics of a dependency and discuss
how they relate to the risk of failure or sabotage from a beacon client
perspective.  

As we describe in detail in~\S\ref{ssec:meth}, it is in general easy to
automatically enumerate all dependencies via language-specific tools, or
simple scripts (see Appendix~\ref{ap:scripts}).
To better assess the relative importance of dependencies for a project,
it would be valuable to determine how much of the code and API of a
dependency is used by the parent project.
But this requires more complex tools, so we restricted ourselves to
identifying dependencies.

A further limitation is when dependencies are not inventoried by the
package manager, but are directly copied into the code tree as source
code or compiled libraries. Specific software is then needed to identify
these.

\paragraph{Version.} 
Outdated versions are indisputably an indicator of potential security issues. 
Indeed, older versions can contain unpatched vulnerabilities, and also have
lower performance and general quality (although recent versions can also
be less stable).
However, some projects might refrain from updating to newer versions
unless the older versions create a security risk: newer versions may be
less stable and include performance or stability regressions.

It is generally easy to automatically determine dependencies versions,
via the package manager or simple scripts.

\paragraph{Known Vulnerabilities.} 
Known vulnerabilities include all the unpatched \emph{known to exist} 
security issues in a project, as published via security advisories, online
articles, and public issue trackers, for example.
This does not include documented security limitations and design
choices.

Of course, relying on a vulnerable component does not mean that the
parent project can be exploited via this vulnerability: the affected
code in the dependency may not be used, or the parent project may not
provide an attack vector to exploit the vulnerability.
As noted in~\cite{PLATE18}:
\begin{quote}
The vast majority (81\%) of vulnerable dependencies may be fixed by
simply updating to a new version, while 1\% of the vulnerable
dependencies in our sample are halted, and therefore, potentially
require a costly mitigation strategy.
\end{quote}

To automate the detection of known vulnerabilities, language-specific
platforms can be used (such as RustSec for Rust, via \code{cargo-audit}.
However, these won't report bugs for which an official advisory wasn't
created, for example, bugs in the issue tracker.

\paragraph{Degree.} 
The set of actual dependencies of a project is not a flat list, but a
graph that includes direct dependencies as well as all
dependencies-of-dependencies.  We call direct dependencies
\emph{first-degree}, dependencies of a direct dependency
\emph{second-degree} ones, and so on.  Note that the same dependency may
occur at different places in the
dependency graph, and under different versions (for example, code
auditors will be familiar with multiple versions of the \code{rand} in
Rust). 

As a rule of thumb, a parent project is
likely to be dependent on a direct dependency than a higher-degree
one (a greater fraction of the code and API are used than for a high-degree
dependency), but this varies a lot.
However, higher-degree dependencies are less visible to developers, less
likely to be noticed by auditors, and thus better targets for backdoors.

Dependency listing tools and scripts usually provide a way to determine
the dependencies of a given degree.

\paragraph{Language.} 
Memory-safe languages are intrinsically safer than (say) C/C++, because
memory-safety (almost) eliminates the whole class of memory corruption bugs,
which are arguably the main cause of exploitable vulnerabilities.

The beacon clients reviewed are written in memory-safe languages, and so are most of their dependencies, but they may still be exposed to memory corruption bugs via dependencies in other languages (Prysm has C++ dependencies, for example), language limitations (Go can SIGSEV\footnote{\url{https://blog.stalkr.net/2015/04/golang-data-races-to-break-memory-safety.html}}, for example), or ``unsafe'' components such as Rust's \code{unsafe} blocks\footnote{\url{https://shnatsel.medium.com/auditing-popular-rust-crates-how-a-one-line-unsafe-has-nearly-ruined-everything-fab2d837ebb1}}.

It is, in general, easy to automatically determine if a project has direct
dependencies in other languages, but it is trickier for dependencies.
Indeed, third-party code in a different language may not be managed via
the language's package manager, or might be directly embedded in the
code as C (e.g. via cgo) or assembly. 

\paragraph{Criticality.} 
It is tempting to categorize dependencies in categories such as 
\begin{itemize}
    \item Critical: Code that performs cryptographic operations or processes attacker-controlled input (deserialization, parsers, etc.).
    \item Non-critical: The rest.
\end{itemize} 
Indeed, a failure of or bug in such critical components is more likely to have more severe consequences for users.
However, when it comes to sabotage and backdoors, the situation is a bit different.
Indeed, maintainers and code auditors will likely pay less attention to non-critical dependencies, such as a package to change the color of a button in the UI.
On the one hand, ultimately all dependencies execute code at the same privilege level, and can potentially (for example) access the filesystem.
On the other hand, a harmful modification in a non-critical component will likely be more visible and obvious than one in a critical component, where a change of a single line or single character may have a dramatic impact.

Identifying critical components generally requires manual review, and is
hard to extend to the whole dependency graph.
In the beacon clients reviewed, critical components include for
example BLS signatures logic (blst and wrappers thereof).

\paragraph{Popularity.} 
The more a project is used and established, the more likely bugs are to
be detected and fixed. A simple indicator of an open-source project's
popularity is its number of GitHub stars (or equivalent rating on other
platforms).
This can be easily collected via GitHub's API.

\paragraph{SSDLC Quality.} 
Does the project have a CI pipeline with extensive testing and code
coverage estimates? Are static analysis tools used? Has the project been
audited by external experts? Are reported issues addressed in a timely
manner? 
These aspects, and others related to secure development lifecycle, are a
major indicator of a project's risk assurance.
We discuss some of these points in~\S\ref{sec:sw}.

\subsection{Methodology}\label{ssec:meth}

This section describes how we collected information about each project's
dependencies, providing reproducible instructions and documenting the
limitations of our approach.




\paragraph{Lighthouse (Rust).} 
Rust's package manager is Cargo and the dependencies are defined in
\path{Cargo.toml} files, wherein the \code{[dependencies]} section lists
the project's dependencies. 
The \path{Cargo.lock} lockfile is generated by Cargo to provide a
deterministic state of the build,
Lighthouse's \path{Cargo.toml} looks like this:

{
\footnotesize
\begin{verbatim}
    [package]
    name = "lighthouse"
    version = "1.3.0"
    authors = ["Sigma Prime <contact@sigmaprime.io>"]
    edition = "2018"
    
    [features]
(...)
    
    [dependencies]
    beacon_node = { "path" = "../beacon_node" }
    tokio = "1.1.0"
    slog = { version = "2.5.2", features = ["max_level_trace"] }
    sloggers = "1.0.1"
    types = { "path" = "../consensus/types" }
    bls = { path = "../crypto/bls" }
(...)
\end{verbatim}
}

Various Cargo extensions help in analyzing dependencies:
\begin{itemize}
    \item \code{cargo-audit}
    utility\footnote{\url{https://github.com/RustSec/rustsec}} reviews
    dependencies for known vulnerabilities recorded in the RustSec
    database~\cite{RUSTSEC}.    

    \item \code{cargo-tree} creates a tree view.

    \item \code{cargo-depgraph}\footnote{\url{https://crates.io/crates/cargo-depgraph/}}
    creates dependency graphs using \code{cargo-metadata} and Graphviz. 

    \item \code{cargo-deps}\footnote{\url{https://crates.io/crates/cargo-deps}} is
    another tool to create graphs, requiring though the manifest to not be
    virtual.

    \item \code{cargo-real-deps}\footnote{\url{https://github.com/Geal/cargo-real-deps}}
    lists dependencies for specific packages in the workspace and
    specific build parameters (which proved useful for Lighthouse, as
    its workspace has a virtual manifest);

    \item \code{cargo-geiger}
    utility\footnote{\url{https://github.com/rust-secure-code/cargo-geiger}}
    reviews the usage of \code{unsafe} blocks.

    \item \code{cargo-udeps} finds which dependencies in \path{Cargo.toml} are unused.

    \item \code{cargo-outdated}\footnote{\url{https://crates.io/crates/cargo-outdated}}
    reveals which components have a newer version available. 

    \item \code{cargo vendor} can vendor all dependencies in a local directory.

\end{itemize}
Not all of these proved applicable to our analysis, and we notably used 
\code{cargo-update} to look at the
\path{Cargo.lock} lockfile and update all the dependencies that have a
higher version than the one defined. Although not exactly the desired
functionality, the output can prove useful to find and count the
outdated dependencies.
Appendix~\ref{ap:scripts:rust} contains the script we used.

\paragraph{Prysm (Go).} 

Go projects list dependencies in the \path{go.mod} file and the command
\code{go list -m all} provides additional information. 
The \code{-u} flag can be used to fetch the latest version available for
each module required is listed.
Some projects might include the legacy method with a \path{vendor/}, but
Prysm uses Go modules.
A project using Go modules can vendor all its dependencies in a local directory using the command \code{go mod vendor}
Prysm's \path{go.mod} looks like this:
{
\footnotesize
\begin{verbatim}
module github.com/prysmaticlabs/prysm

go 1.16

require (
        contrib.go.opencensus.io/exporter/jaeger v0.2.1
        github.com/StackExchange/wmi v0.0.0-20210224194228-fe8f1750fd46 // indirect
        github.com/allegro/bigcache v1.2.1 // indirect
        github.com/aristanetworks/goarista v0.0.0-20200521140103-6c3304613b30
        github.com/bazelbuild/buildtools v0.0.0-20200528175155-f4e8394f069d
(...)
\end{verbatim}
}

The \path{go.mod} file includes only the direct dependencies and some
indirect (with the suffix \code{// indirect}), when those are not listed
in the \path{go.mod} file of the direct dependency or if the direct
dependency does not have a \path{go.mod} file. 
Also, any dependency that is not imported in the module's source files
is marked as \code{// indirect}. 
To collect all Prysm dependencies, we also used the
\code{go-mod-outdated}\footnote{\url{https://github.com/psampaz/go-mod-outdated}}
utility, which produces a Markdown table view of the \code{go list -u -m
-json all} listing all dependencies of a Go project and their available minor and patch updates. 
Our script to collect information on Prysm
Appendix~\ref{ap:scripts:prysm} uses that tool.

The \path{nancy}
utility\footnote{\url{https://github.com/sonatype-nexus-community/nancy}}
by Sonatype can then be used to automatically search for entries in
vulnerability databases, from the output of \code{go list -json -m all}.

Furthermore, as noted in 
\path{DEPENDENCIES.md}\footnote{\url{https://github.com/prysmaticlabs/prysm/blob/develop/DEPENDENCIES.md}},
``Prysm is go project with many complicated dependencies, including
some c++ based libraries.''
The latter are managed via Bazel, including precompiled libraries for
convenience, for which the source code is not included.
Bazel is also used to integrate local patches to Prysm's dependencies,
``to make a small change (...) for ease of use in Prysm''.


\paragraph{Teku (Java).}
The two main package managers in Java are Gradle and Maven. 
Teku uses Gradle, which describes the dependencies in the \path{build.gradle} configuration file. 
A list of dependencies (direct and indirect) can be obtained via \code{gradle -q dependencies}~\cite{GRADLE} which provides a good visualization of the dependency tree. Also, a useful tool is the build scans for the Gradle feature provided in \url{https://scans.gradle.com/} as well as developers can find information about the artifacts and their versions in \url{https://mvnrepository.com/}. Teku's \path{teku/build.gradle} looks like this:

{
\footnotesize
\begin{verbatim}
dependencies {
    implementation 'com.google.guava:guava'
    implementation 'org.apache.commons:commons-lang3'
    implementation 'org.apache.logging.log4j:log4j-api'
    runtimeOnly 'org.apache.logging.log4j:log4j-core'
    runtimeOnly 'org.apache.logging.log4j:log4j-slf4j-impl'
    testImplementation 'org.apache.tuweni:tuweni-junit'
    testImplementation 'org.assertj:assertj-core'
    testImplementation 'org.mockito:mockito-core'
    testImplementation 'org.junit.jupiter:junit-jupiter-api'
    testImplementation 'org.junit.jupiter:junit-jupiter-params'
    testRuntimeOnly testFixtures(project(':infrastructure:logging'))
    testRuntimeOnly 'org.junit.jupiter:junit-jupiter-engine'
    testFixturesImplementation 'org.assertj:assertj-core'
(...)
  }
\end{verbatim}
}

Note that Gradle, unlike Cargo, resolves ``version conflicts'' when two
or more components use different versions of the same dependency:
Gradle will choose the most recent version of all versions appearing in
the dependency graph, as described in Gradle's
documentation\footnote{\url{https://docs.gradle.org/current/userguide/dependency_resolution.html}}.

Additionally, each dependency is characterized by a
\textit{Configuration} that defines its
scope\footnote{\url{https://docs.gradle.org/current/userguide/dependency_resolution.html}}
(for example, for runtime, testing, building).
Each configuration has a specific configurable name, however, many Gradle plugins have pre-defined configurations. 
This is the case for Teku. 
Based on their scope, we decided to count only the direct dependencies with \code{implementation} configuration.

Our script in Appendix~\ref{ap:scripts:teku} finds the number of direct
dependencies, the number of total dependencies and their maximal degree.

\paragraph{Nimbus (Nim).}

The Nim language has a package manager called Nimble, which lists
dependencies in \path{.nimble} files~\footnote{\url{https://github.com/nim-lang/nimble}}.
However, Nimbus does not use Nimble but instead lists dependencies as git submodules in a \path{vendor/} directory\footnote{\url{https://github.com/status-im/nimbus-eth2/tree/stable/vendor}}.
These include:
{
    \footnotesize
\begin{verbatim}
NimYAML @ ca82b5e
asynctools @ c478bb7
eth2-testnets @ 5b4e327
jswebsockets @ ff0ceec
karax @ 32de202
news @ 002b21b
nim-bearssl @ 0a7401a
(...)
\end{verbatim}
}

The \path{vendor} folder approach avoids incompatibilities in the case that two projects use a different version. 
For most dependencies, the version integrated is not a release but a
certain commit, which may introduce extra risks (non-release versions
are likely to be less stable) if it is not frequently updated.
In comparison, the Nimble package manager always fetches and installs
the latest release of a repository or the latest commit if there are no
releases or there is the \code{\#head} as suffix.

Our script in Appendix~\ref{ap:scripts:nim} lists the direct dependencies of a project through the \code{.nimble} file. 
However, this was not used for Nimbus, for which we just listed the
content of \path{vendor/}. 
As the direct dependencies were few and the max degree of them was
three, the total dependencies were counted by inspecting each
component's \code{.nimble} file.

\subsection{Dependencies Review}\label{sec:deps:res}

This section describes the results of our dependencies review for the four beacon clients. 


Table~\ref{tab:metrics} describes the risk indicators to the four beacon
clients reviewed.
Below we comment further on these results, and how we observed them
evolve over time, which provides insights into the dependency management
of the projects.

\begin{table}    
    \centering
    \begin{tabular}{lllll}
        \toprule
        \textbf{Metric} & \textbf{Lighthouse} & \textbf{Nimbus} & \textbf{Prysm} & \textbf{Teku} \\
        \midrule
        Language & Rust & Nim & Go & Java \\
        GitHub Stars & 1.2k & 212 & 2.2k & 257 \\
        Direct dependencies & 121 & 43 & 93 & 48 \\
        Total dependencies & 440 & 56 & 665 & 230 \\
        Max degree of dependencies & 15 & 3 & 13 & 13 \\
        Outdated versions & 59 & 0 & 353 & N/A \\
        Vulnerable versions & 5 & 0 & 5 & 18 \\
        CVEs & 6 & 0 & 11 & 23 \\
        Last commit & 10/06/21 & 05/08/21 & 10/08/21 & 06/08/21 \\
        Last release & 10/06/21 & 05/08/21 &  03/08/21 & 28/07/21 \\
        Open issues & 100 & 150 & 97 & 81 \\
        Closed issues & 816 & 514 & 1999 & 1215 \\
        \bottomrule
    \end{tabular}
    \caption{Overview of the risk metrics, as of 20210810.}\label{tab:metrics}
\end{table}

\paragraph{Lighthouse (Rust).} 

As of 20210730, Lighthouse had 5 dependencies with vulnerable versions.
\begin{itemize}
    \item \path{libsecp256k1} v0.3.5 was affected by \href{https://nvd.nist.gov/vuln/detail/CVE-2021-38195}{CVE-2021-38195}
    \item \path{prost-types} v0.7.0 was affected by \href{https://nvd.nist.gov/vuln/detail/CVE-2021-38192}{CVE-2021-38192}
    \item \path{tokio} v0.3.7 and v1.5.0 were affected by \href{https://nvd.nist.gov/vuln/detail/CVE-2021-38191}{CVE-2021-38191}
    \item \path{crossbeam-deque} v0.8.0 is affected by \href{https://nvd.nist.gov/vuln/detail/CVE-2021-32810}{CVE-2021-32810}
    \item \path{hyper} v0.13.10 and v0.14.7 are affected by \href{https://nvd.nist.gov/vuln/detail/CVE-2021-32714}{CVE-2021-32714}
    \item \path{hyper} v0.13.10 and v0.14.7 are affected by \href{https://nvd.nist.gov/vuln/detail/CVE-2021-32715}{CVE-2021-32715}
\end{itemize}
Except for the vulnerable crates, there were reported 13 unmaintained crates,
the 9 of them were merged to different crates, 3 were totally
unmaintained (one for over three years) and one was deprecated. 
However, running the same analysis one month later gave different results.

As of 20210824, Lighthouse had only 1 dependency with known vulnerabilities:
\begin{itemize}
	\item \path{openssl-src} version \path{111.15.0+1.1.1k} is affected by \href{https://nvd.nist.gov/vuln/detail/CVE-2021-3711}{CVE-2021-3711} and \href{https://nvd.nist.gov/vuln/detail/CVE-2021-3712}{CVE-2021-3712}
\end{itemize}
This time, the tool reported only 2 unmaintained crates. Checking only
after few days, developers had patched the vulnerable component the next
day of the CVE release in the Rust vulnerability database and the report
of the \code{cargo-audit} found no vulnerabilities.  This suggests that
the maintainers regularly review their
dependencies and upgrade the vulnerable components to their higher
versions which fix the security bugs. After searching several pull
requests for Lighthouse, it was found that they make use of bors
bot~\cite{BORS} to run automated CI tasks. 
For Lighthouse, this includes both \code{cargo-audit} and \code{cargo-udeps}. 

\begin{figure}
	\centering
	\includegraphics[width=1.0\textwidth]{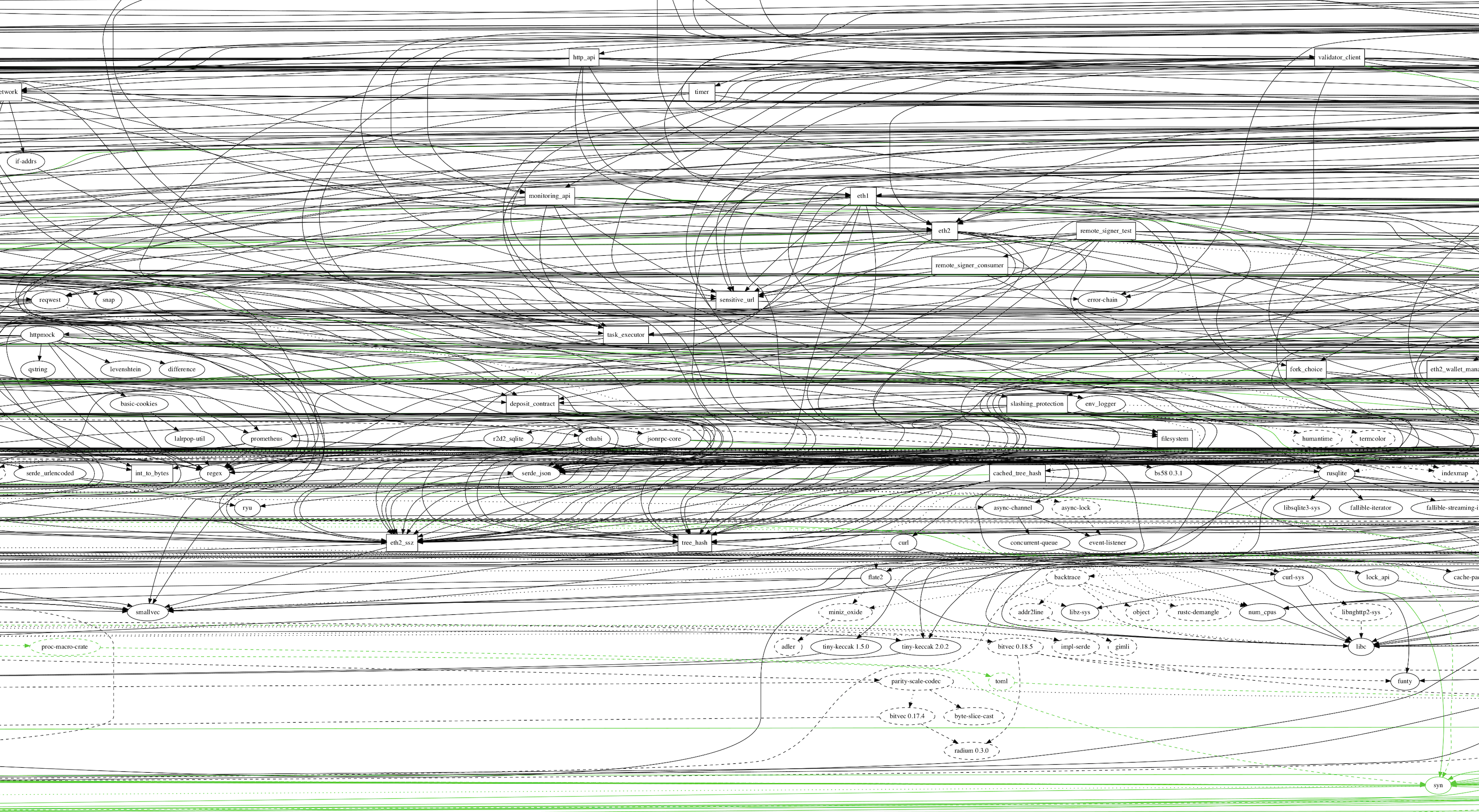}
	\caption{Lighthouse dependencies graph (excerpt).}
	\label{fig:deps:rustgraph}
 \end{figure}

As of 20210825, after running the \code{cargo-update}, Lighthouse was found to have 59 of the total dependencies outdated.
Figure~\ref{fig:deps:rustgraph} depicts a small part of the graph generated by using the \code{cargo depgraph}. 
The fact that almost nothing is discernible confirms the complexity of
dependencies in such big projects and how difficult it is to monitor
them and eliminate the attack vectors through them. 
Finally, the max degree of the dependencies was 15, which was found by
observing the different alignments as printed by the \code{cargo-tree}
(see the script in Appendix~\ref{ap:scripts:rust}).

\paragraph{Prysm (Go).} 

\begin{figure}
	\centering
	\includegraphics[width=1.0\textwidth]{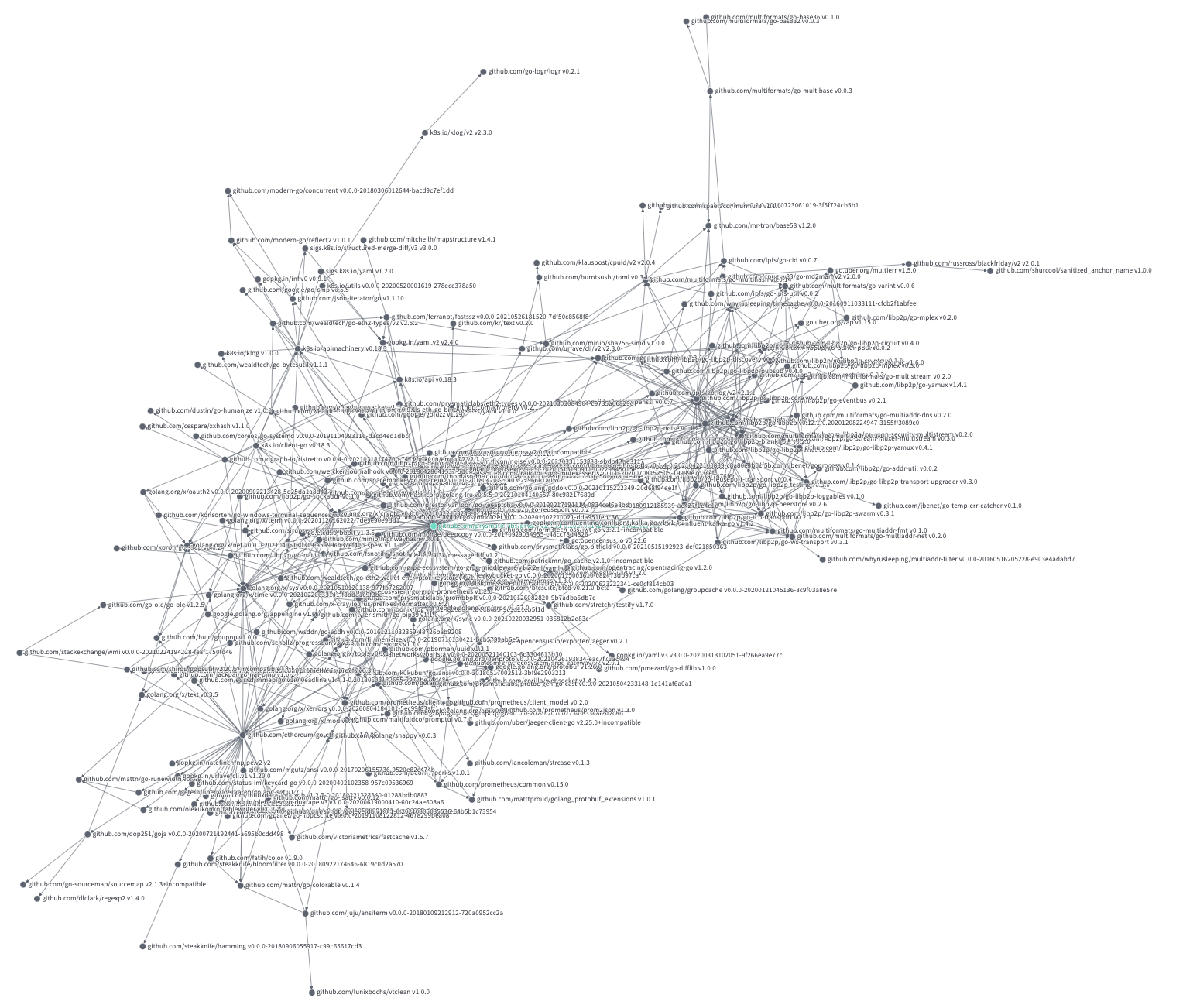}
	\caption{Prysm dependencies graph, generated using \url{https://deps.dev/}.}
	\label{fig:deps:prysmgraph}
\end{figure}

Figure~\ref{fig:deps:prysmgraph} shows Prysm's dependency graph.
As of 20210731, Prysm had 93  Go direct dependencies, 45 out of which were outdated. 
In total, it had 599 dependencies, 332 out of which were outdated. 
As of 20210831, a month later, the state was almost the same. 43 out of
91 direct dependencies were outdated, as well as 369 out of the 658
total dependencies. 
Most of the outdated components were the same.

As of 20210811, the \path{nancy sleuth} utility reported 2 vulnerable
components.  As of 20210605, OWASP's Dependency-Check reported 37
vulnerable dependencies out of the 596 scanned.  However, after careful
manual inspection, 32 of the 37 were found to be false positives.
We also matched vulnerabilities against the Snyk\footnote{\url{https://snyk.io}}, and OSV databases. 
Table~\ref{tab:prysmcves} reports the CVEs found in those 5 components.


\begin{table}    
    \centering
    \begin{tabular}{llll}
        \toprule
        \textbf{Dependency} & \textbf{Version} & \textbf{Versions affected} & \textbf{CVE} \\
        \midrule
        \path{coreos/etcd} & v3.3.13 & prior to v3.3.23 & \href{https://nvd.nist.gov/vuln/detail/CVE-2020-15114}{CVE-2020-15114} \\
        \path{coreos/etcd} & v3.3.13 & prior to v3.3.23 & \href{https://nvd.nist.gov/vuln/detail/CVE-2020-15113}{CVE-2020-15113} \\
        \path{coreos/etcd} & v3.3.13 & prior to v3.3.23 & \href{https://nvd.nist.gov/vuln/detail/CVE-2020-15112}{CVE-2020-15112} \\
        \path{coreos/etcd} & v3.3.13 & prior to v3.3.23 & \href{https://nvd.nist.gov/vuln/detail/CVE-2020-15106}{CVE-2020-15106} \\
        \path{dgrijalva/jwt-go} & v3.2.0 & prior to v4.0.0-preview1 & \href{https://nvd.nist.gov/vuln/detail/CVE-2020-26160}{CVE-2020-26160} \\
        \path{hashicorp/consul/api} & v1.3.0 & $\geq$ v1.2.0 prior to v1.6.6 & \href{https://nvd.nist.gov/vuln/detail/CVE-2020-13250}{CVE-2020-13250} \\
        \path{hashicorp/consul/api} & v1.3.0 & prior to v1.6.2 & \href{https://nvd.nist.gov/vuln/detail/CVE-2020-7219}{CVE-2020-7219} \\
        \path{nats-io/nats-server/v2} & v2.1.2 & prior to v2.2.0 & \href{https://nvd.nist.gov/vuln/detail/CVE-2020-28466}{CVE-2020-28466} \\
        \path{nats-io/nats-server/v2} & v2.1.2 & prior to v2.1.9 & \href{https://nvd.nist.gov/vuln/detail/CVE-2020-26521}{CVE-2020-26521} \\
        \path{nats-io/nats-server/v2} & v2.1.2 & prior to v2.1.9 & \href{https://nvd.nist.gov/vuln/detail/CVE-2020-26892}{CVE-2020-26892} \\
        \path{nats-io/nats-server/v2} & v2.1.2 & prior to v2.2.0 & \href{https://nvd.nist.gov/vuln/detail/CVE-2021-3127}{CVE-2021-3127} \\
        \path{nats-io/jwt} & v0.3.2 & prior to v2.0.1 & \href{https://nvd.nist.gov/vuln/detail/CVE-2021-3127}{CVE-2021-3127} \\
        \bottomrule
    \end{tabular}
    \caption{Overview of CVEs found in Prysm dependencies.}\label{tab:prysmcves}
\end{table}

Note that these results were obtained on the latest commit of the
development branch at the time of the test.
It is possible that, prior to issuing a new release, the Prysm
developers update all dependencies, but we could not verify it.

\paragraph{Teku (Java).} 

As of 20210727, Teku had 48 direct dependencies, counting only the
components with the configuration \code{implementation}.
In total, 230 dependencies are used in the Teku development, of which 18
included known vulnerabilities, as collected from the Snyk database:
\begin{itemize}
\item \path{org.webjars:swagger-ui:3.25.2} is affected by \href{https://snyk.io/vuln/SNYK-JAVA-ORGWEBJARS-575003}{SNYK-JAVA-ORGWEBJARS-575003}
\item \path{org.web3j:rlp:4.6.2} is affected by \href{https://snyk.io/vuln/SNYK-JAVA-ORGWEB3J-598385}{SNYK-JAVA-ORGWEB3J-598385}
\item \path{org.mozilla:rhino:1.7R4} is affected by \href{https://snyk.io/vuln/SNYK-JAVA-ORGMOZILLA-1314295}{SNYK-JAVA-ORGMOZILLA-1314295}
\item \path{org.java-websocket:Java-WebSocket:1.3.8} is affected by \href{https://nvd.nist.gov/vuln/detail/CVE-2020-11050}{CVE-2020-11050}
\item \path{org.eclipse.jetty:jetty-webapp:9.4.31.v20200723} is affected by \href{https://nvd.nist.gov/vuln/detail/cve-2020-27216}{CVE-2020-27216}
\item \path{org.eclipse.jetty:jetty-servlet:9.4.31.v20200723} is affected by \href{https://nvd.nist.gov/vuln/detail/CVE-2021-28169}{CVE-2021-28169}
\item \path{org.eclipse.jetty:jetty-server:9.4.31.v20200723} is affected by \href{https://nvd.nist.gov/vuln/detail/CVE-2021-34428}{CVE-2021-34428}, \href{https://nvd.nist.gov/vuln/detail/CVE-2020-27218}{CVE-2020-27218} and \href{https://nvd.nist.gov/vuln/detail/CVE-2020-27223}{CVE-2020-27223}
\item \path{org.eclipse.jetty:jetty-io:9.4.31.v20200723} is affected by \href{https://nvd.nist.gov/vuln/detail/CVE-2021-28165}{CVE-2021-28165}
\item \path{org.bouncycastle:bcprov-jdk15on:1.66} is affected by \href{https://nvd.nist.gov/vuln/detail/cve-2020-28052}{CVE-2020-28052}
\item \path{org.apache.commons:commons-compress:1.20} is affected by \href{https://nvd.nist.gov/vuln/detail/CVE-2021-35516}{CVE-2021-35516}, \href{https://nvd.nist.gov/vuln/detail/CVE-2021-35517}{CVE-2021-35517}, \href{https://nvd.nist.gov/vuln/detail/CVE-2021-36090}{CVE-2021-36090} and \href{https://nvd.nist.gov/vuln/detail/CVE-2021-35515}{CVE-2021-35515}
\item \path{io.vertx:vertx-core:3.9.1} is affected by \href{https://nvd.nist.gov/vuln/detail/CVE-2019-17640}{CVE-2019-17640}
\item \path{io.netty:netty-transport:4.1.56.Final} is affected by \href{https://nvd.nist.gov/vuln/detail/CVE-2021-21290}{CVE-2021-21290}
\item \path{io.netty:netty-handler:4.1.51.Final} is affected by \href{https://nvd.nist.gov/vuln/detail/CVE-2021-21290}{CVE-2021-21290} and an \href{https://snyk.io/vuln/SNYK-JAVA-IONETTY-1042268}{SNYK-JAVA-IONETTY-1042268}
\item \path{io.netty:netty-common:4.1.56.Final} is affected by \href{https://nvd.nist.gov/vuln/detail/CVE-2021-21290}{CVE-2021-21290}
\item \path{io.netty:netty-codec-http:4.1.51.Final} is affected by \href{https://nvd.nist.gov/vuln/detail/CVE-2021-21290}{CVE-2021-21290}, \href{https://nvd.nist.gov/vuln/detail/CVE-2021-21295}{CVE-2021-21295} and an \href{https://snyk.io/vuln/SNYK-JAVA-IONETTY-1020439}{SNYK-JAVA-IONETTY-1020439}
\item \path{io.netty:netty-codec-http2:4.1.51.Final} is affected by \href{https://nvd.nist.gov/vuln/detail/CVE-2021-21295}{CVE-2021-21295} and \href{https://nvd.nist.gov/vuln/detail/CVE-2021-21409}{CVE-2021-21409}
\item \path{commons-io:commons-io:2.6} is affected by \href{https://nvd.nist.gov/vuln/detail/CVE-2021-29425}{CVE-2021-29425}
\item \path{com.google.guava:guava:29.0-jre} is affected by \href{https://nvd.nist.gov/vuln/detail/CVE-2020-8908}{CVE-2020-8908}
\end{itemize}

\paragraph{Nimbus (Nim)).} 

As of 20210802, Nimbus had 43 direct dependencies, as included in the
\path{vendor} folder. 
In total, the project used 56 dependencies.
After a manual investigation in the NVD and the Snyk databases, no
known vulnerability was found to affect the dependencies.
Moreover, no dependency was found to be outdated. 
As of 20210902, one month later, the results were the same. 
Having neither vulnerable nor
outdated dependencies seems ideal for the security of a project.
The max dependency degree was only three.

This lower ``vulnerability surface'' seems encouraging and a positive
aspect of Nimbus.
However, the Nim language is much less established than Go, Java, or
Rust, and the security of the language and its runtime is arguably
underanalyzed.

\subsection{Discussion and Recommendations}


The two main beacon client projects, Lighthouse and Prysm, each depend
on hundreds of third-party projects (cf. Table~\ref{tab:metrics}).
These projects appear to pay attention to security in
their dependencies, as only 5 out of hundreds had known vulnerabilities.
Lighthouse seems to be more diligent, with CI checks and more regular
updates.

All projects, as all successful open-source projects, are flooded with
issues in their GitHub Issues tracker, and Prysm seems to be the more
effective at processing and closing issues.
For an attacker, open issues and the associated discussions can be
goldmines of information, especially when projects don't document
how to properly report security issues: Lighthouse and Prysm provide a
dedicated contact and a PGP key; Teku provides a dedicated but no PGP
key; Nimbus does not describe any process.

From an attacker's perspective, Prysm appears to be the most attractive
target: the most widely used, the highest number of dependencies, and
``lesser'' secure SDLC procedures than Lighthouse.
Many direct dependencies of Prysm are from more trustworthy
sources (from \path{golang.org/x/}, \path{k8s.io/},
\path{github.com/google/}, for example), however, it also uses some wallet
encryption project with 1 GitHub star
(\href{https://github.com/wealdtech/go-eth2-wallet-encryptor-keystorev4}{wealdtech/go-eth2-wallet-encryptor-keystorev4})
among its direct dependencies.
That said, as discussed in~\S{\ref{sec:deps:metrics} under
``Criticality'', planning sabotage of such obviously security-critical
projects might not be the best approach for an attacker.


Beacon clients being critical components of the Ethereum environment,
attackers may devote significant resources to compromising them.
Although the larger number of clients---and thus amount of code---makes
for a wider attack surface, and more bugs, we argue that from a risk
management perspective some diversity of beacon clients is beneficial.
As Lighthouse developers
argued\footnote{\url{https://lighthouse.sigmaprime.io/switch-to-lighthouse.html}},
``client diversity'' prevents coordinated failure of a large part of the
network, as it happened after a Prysm bug.

Therefore, rather than encouraging the use of only Lighthouse
and Prysm, we would more strongly encourage 1) the deployment of
nodes using all four beacon clients, and 2) all beacon clients to
adopt more mature security practices to prevent counterparty risks and
supply chain attacks, including: 

To reduce the risk of supply chain attacks (via known or maliciously
introduced vulnerabilities), beacon clients and  

\begin{itemize}
    \item Document and enforce a dependency management policy, notably
    setting criteria on the acceptable dependencies.

    \item Integrate automatic checks for dependencies versions and
    vulnerability as part of the CI pipeline, using language-specific
    tools and relevant GitHub features (Actions,
    dependabot\footnote{\url{https://github.blog/2021-04-29-goodbye-dependabot-preview-hello-dependabot/}},
    etc.).

    \item Keep track of all the dependencies (direct and indirect) in a
    ``software bill of materials'', to have visibility on the project's
    liabilities, and help in monitoring external projects.

    \item Minimize the number of third-party dependencies

    \item When security audits are organized, consider including the
    most critical external dependencies in the scope.

    \item Use multiple sources and databases of vulnerabilities, and
    don't blindly trust composition tools' results (check for false
    positives).

    \item Encourage or enforce signed commits for all contributors of
    the project (``hypocrite commits'' may not only be in dependencies).
\end{itemize}

\section{Network Fingerprinting}\label{sec:fngp}

Although nodes and validators addresses are by definition public, it
doesn't mean that all client's services are public too and can be
exposed to the internet.
For example, the beacon node API, as discussed in~\S\ref{sec:api}.
Moreover, exposing metadata such as a client type and version, operating
system details, could be leveraged by an attacker.

In this section, we thus analyze the visibility and discoverability of
beacon clients, against internet scans, and through host the host search
engines Censys, FOFA, and Shodan.

\subsection{Exposure Overview}

Using the methodology from~\cite{SDWAN}, we investigated the resistance
of the beacon clients interfaces to fingerprinting techniques.
Fingerprints may for example be used by an attacker to scan the internet
and look for:
\begin{itemize}
  \item Clients running a given beacon client software (e.g., Nimbus),
  or a specific version (range) thereof.
  \item Clients with an insecure configuration (e.g., the beacon node API exposed to the internet).
  \item Hosts with a given operating system version. 
\end{itemize}
In general, Ethereum services are not trivially discoverable via search
engines or hosts databases, because Ethereum relies on
non-standard protocols and APIs.
However, most services will leak some metadata that will permit their
identification. 
In particular, the Prysm's web interface, which can be identified
via favicon hash, HTML page title, and CSS keywords. 
Appendix~\ref{ap:fingerprints} includes some search engine queries we
determined for all the beacon clients. 


We found the most fingerprinting-friendly components to be:
\begin{itemize}
  \item \code{/eth/v1/node/version} Ethereum API endpoint: Retrieves the beacon node information about its implementation in a format similar to an HTTP User-Agent header.
  \item \code{/eth/v1/node/peers} Ethereum API endpoint: Retrieves data about the node's network peers, returning all peers.
  \item \code{/lighthouse/peers} and \code{/lighthouse/peers/connected} Lighthouse-specific API endpoints\footnote{\url{https://lighthouse-book.sigmaprime.io/api-lighthouse.html}}: Provide agent name and internal IP addresses for each listed peer.
  \item libp2p agent version: a free-form characteristic string, identifying the implementation of the peer\footnote{\url{https://github.com/libp2p/specs/blob/master/identify/README.md\#agentversion}}.
\end{itemize}

Note that a lot of information is available directly from the nodes:
Lighthouse API's endpoint \code{/lighthouse/peers} provides information about a peers'  OS and client version retrieved from an agent version:
{
\footnotesize
\begin{verbatim}

"client":{
    "kind":"Teku",
    "version":"v21.3.2",
    "os_version":"linux-x86_64",
    "protocol_version":"ipfs/0.1.0",
    "agent_string":"teku/teku/v21.3.2/linux-x86_64/oracle_openjdk-java-15"
}

\end{verbatim}
}

All clients provide different information in that field:
\begin{itemize}
  \item Lighthouse: client name, client version, OS version
  \item Teku: client name, client version, OS version
  \item Nimbus: client name
  \item Prysm: client name, client version
\end{itemize}
It is unclear whether Nimbus' information is less verbose on purpose.

Also, the \code{/lighthouse/peers/connected} endpoint discloses IP-addresses of internal networks, for example:
{
\footnotesize
\begin{verbatim}
"peer_id":"26Uiu2HBmG7bMnsJPWE2t3PQNcgihtrsYr1kXtrtzEfh7QFvCwhEw",
"peer_info":{
    "_status":"Healthy",
    "client":{
        "kind":"Lighthouse",
        "version":"v1.4.0-3b600ac",
        "os_version":"x86_64-linux",
        "protocol_version":"lighthouse/libp2p",
        "agent_string":"Lighthouse/v1.4.0-3b600ac/x86_64-linux"
    },
    "listening_addresses":[
        "/ip4/54.169.172.126/tcp/9000",
        "/ip4/127.0.0.1/tcp/9000",
        "/ip4/10.0.48.248/tcp/9000",
        "/ip4/192.168.0.104/tcp/9000"
    ]
}
\end{verbatim}
}

Disclosing the private IP addresses does not seem necessary, and can
help an attacker, for example in SSRF-like attacks, as discussed in
\S\ref{ssec:apiauth}.

\subsection{Clients Fingerprints Examples}

For each client, we give an example of fingerprint:

\paragraph{Lighthouse.}
Instances can be found from endpoints served
on port 5200\footnote{\url{https://lighthouse-book.sigmaprime.io/docker.html}},
such as /eth/v1/node/version. Here's an example of a response from a Lighthouse node:
{
\footnotesize
\begin{verbatim}

{"data":{"version":"Lighthouse/v1.4.0-3b600ac/x86\_64-linux"}}''

\end{verbatim}
}

\paragraph{Nimbus.}
Instances may be found from enpoints served on
 port 5052\footnote{\url{https://github.com/status-im/nimbus-eth2/blob/stable/beacon\_chain/spec/network.nim\#L33}},
 by JSON RPC ports 9190 and sometimes 9091\footnote{\url{https://nimbus.guide/api.html}},
 and in rare cases also by metrics on port 8008\footnote{\url{https://github.com/status-im/nimbus-eth2/blob/stable/Jenkinsfile\#L51}}.
Note that we couldn't discover Nimbus nodes with a public REST API.
A reason might be that Nimbus' REST API is in beta and disabled by
default, accessible locally only\footnote{\url{https://nimbus.guide/rest-api.html}}.

\paragraph{Prysm.}
Instances can be found from their web interface, served by default
on port 7500\footnote{\url{https://github.com/prysmaticlabs/prysm-web-ui/blob/master/src/environments/environment.ts\#L8}}.
They can be also identified by the HTTP title "PrysmWebUi".

\paragraph{Teku.}
Instances can be found from endpoints served 
on port 5051\footnote{\url{https://docs.teku.consensys.net/en/latest/HowTo/Get-Started/Installation-Options/Run-Docker-Image/}}.
Sometimes, Teku's API can be found on port 5052. Here's an example of a response from a Teku node:
{
\footnotesize
\begin{verbatim}

{"data":{"version":"teku/v20.11.1/windows-x86\_64/oracle-java-15"}}

\end{verbatim}
}

Moreover, some Teku instances include Prometheus metrics on port 8008, which
also exposes version and other information, for example: {
\footnotesize
\begin{verbatim}

# HELP beacon_teku_version Teku version in use
# TYPE beacon_teku_version counter
beacon_teku_version{version="21.6.1+4-gca9294a",} 1.0

\end{verbatim}
}

\subsection{Nodes Discovery}

We used the following base method to discover beacon nodes:
\begin{enumerate}
\item Take a list of the nodes found using the search engines
\item For each node from the list, query its API to retrieve the node's peers
\item Add the new nodes into the list
\item Scan potentially exposed network ports of the node
\item If possible, get a client name and version of the node using \code{/eth/v1/node/version} endpoint
\item Go to step 1 with a new extended list of nodes and repeat
\end{enumerate}

We discovered about 12\,000 nodes, using only the method described above without any P2P mechanisms and libraries.
All found nodes are depicted on Figure~\ref{fig:fngp:map}.
This map can be seen as an indicator of regional usage at a given point in time, but should not be seen as reliable evidence of the distribution of each client's usage,
because of the obvious biases (fingerprints and scans reliability, exposure of the service).

\begin{figure}
	\centering
	\includegraphics[width=1.0\textwidth]{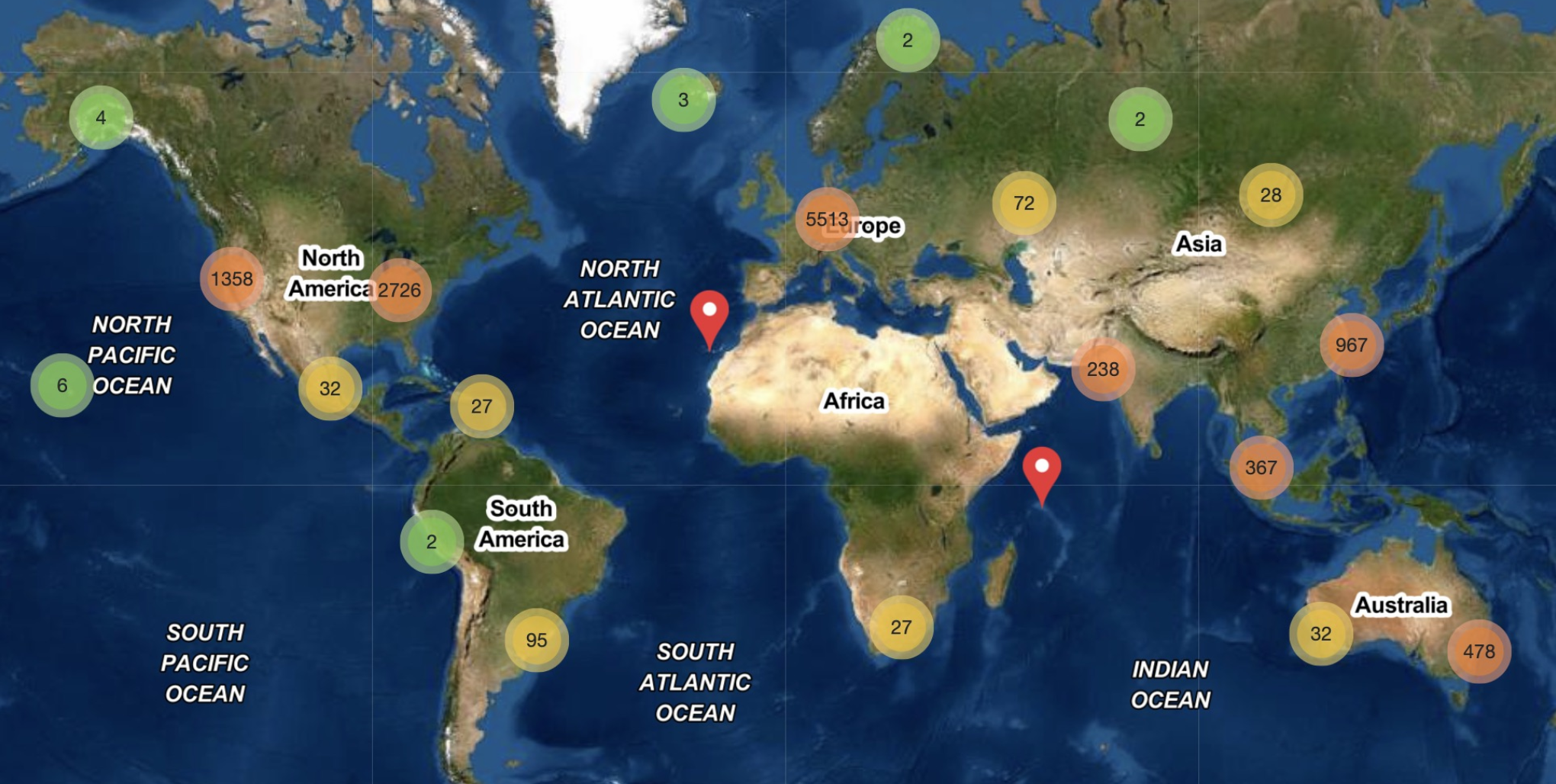}
	\caption{Geolocation of beacon clients identified through fingerprinting.}
	\label{fig:fngp:map}
 \end{figure}

\section*{Acknowledgements}

We would like to thank the Ethereum Foundation for supporting this
project, and Lúcás Meier for his feedback.

\addcontentsline{toc}{section}{Acknowledgements}

\bibliographystyle{abbrv}

\bibliography{bib}

\begin{thebibliography}{10}

\bibitem{BORS}
{BORS} documentation.
\newblock \url{https://bors.tech/documentation/getting-started/}.

\bibitem{eip3076}
{EIP}-3076: Slashing protection interchange format.
\newblock \url{https://eips.ethereum.org/EIPS/eip-3076}.

\bibitem{eth2net}
Ethereum 2.0 networking specification.
\newblock
  \url{https://github.com/ethereum/eth2.0-specs/blob/dev/specs/phase0/p2p-interface.md}.

\bibitem{eth2api}
Ethereum 2.0 specification.
\newblock \url{https://github.com/ethereum/eth2.0-apis}.

\bibitem{eth2spec}
Ethereum 2.0 specification.
\newblock \url{https://github.com/ethereum/eth2.0-specs}.

\bibitem{GITOCTO2020}
{GitHub {Octoverse} security report 2020}.
\newblock
  \url{https://octoverse.github.com/static/github-octoverse-2020-security-report.pdf}.

\bibitem{libp2p-impl}
Introduction to libp2p. {Lesson} 5.
\newblock \url{https://proto.school/introduction-to-libp2p/05}.

\bibitem{libp2p}
Libp2p specification.
\newblock \url{https://github.com/libp2p/specs}.

\bibitem{LhFU}
Lighthouse fuzzing update.
\newblock \url{https://lighthouse.sigmaprime.io/fuzzing-01.html}.

\bibitem{LhSC}
Lighthouse security considerations.
\newblock \url{https://lighthouse.sigmaprime.io/fuzzing-lighthouse.html}.

\bibitem{Lh26}
Lighthouse update \#26.
\newblock \url{https://lighthouse.sigmaprime.io/update-26.html}.

\bibitem{Lh30}
Lighthouse update \#30.
\newblock \url{https://lighthouse.sigmaprime.io/update-26.html}.

\bibitem{VC}
{Lighthouse Validator Client API}.
\newblock \url{https://lighthouse-book.sigmaprime.io/api-vc.html}.

\bibitem{NimRFP}
Nimbus {ETH2.0} security audit request for proposal.
\newblock
  \url{https://our.status.im/nimbus-eth2-0-security-audit-request-for-proposal/}.

\bibitem{NimUp}
Nimbus update: September 11th.
\newblock \url{https://our.status.im/nimbus-update-september-11th/}.

\bibitem{discv5}
Node discovery protocol version 5.
\newblock
  \url{https://github.com/ethereum/devp2p/blob/master/discv5/discv5.md}.

\bibitem{libp2p-noise}
{noise-libp2p - Secure Channel Handshake}.
\newblock \url{https://github.com/libp2p/specs/tree/master/noise}.

\bibitem{noise}
Noise protocol framework.
\newblock \url{http://www.noiseprotocol.org/noise.html}.

\bibitem{RUSTSEC}
The {Rust Security Advisory Database}.
\newblock \url{https://rustsec.org/}.

\bibitem{Nhb}
{The Auditors Handbook to {Nimbus} Beacon Chain}.
\newblock \url{https://nimbus.guide/auditors-book/}.

\bibitem{weaksub2}
A.~Asgaonkar.
\newblock Weak subjectivity in {Eth2.0}.
\newblock \url{https://notes.ethereum.org/@adiasg/weak-subjectvity-eth2}.

\bibitem{TMC}
J.-P. Aumasson.
\newblock Too much crypto.
\newblock \url{https://eprint.iacr.org/2019/1492}.

\bibitem{BLSfv}
J.-P. Aumasson, Q.~T.~M. Nguyen, and A.~Sanso.
\newblock Security of {BLS} fast verification.
\newblock \url{https://hackmd.io/TODO}.

\bibitem{BLScurve}
P.~S. L.~M. Barreto, B.~Lynn, and M.~Scott.
\newblock Constructing elliptic curves with prescribed embedding degrees.
\newblock \url{https://eprint.iacr.org/2002/088}.

\bibitem{BDN18}
D.~Boneh, M.~Drijvers, and G.~Neven.
\newblock Compact multi-signatures for smaller blockchains.
\newblock \url{https://eprint.iacr.org/2018/483}.

\bibitem{BDN18b}
D.~Boneh, M.~Drijvers, and G.~Neven.
\newblock {BLS Multi-Signatures With Public-Key Aggregation}, 2018.
\newblock \url{https://crypto.stanford.edu/~dabo/pubs/papers/BLSmultisig.html}.

\bibitem{BLSIE}
D.~Boneh, S.~Gorbunov, R.~Wahby, H.~Wee, and Z.~Zhang.
\newblock {BLS} signatures.
\newblock
  \url{https://datatracker.ietf.org/doc/draft-irtf-cfrg-bls-signature/}.

\bibitem{BLS01}
D.~Boneh, B.~Lynn, and H.~Shacham.
\newblock {Short Signatures from the {Weil} Pairing}.
\newblock In {\em ASIACRYPT}, 2001.
\newblock \url{https://www.iacr.org/archive/asiacrypt2001/22480516.pdf}.

\bibitem{BLS12381}
S.~Bowe.
\newblock {BLS12-381}: New {zk-SNARK} elliptic curve construction.
\newblock \url{https://electriccoin.co/blog/new-snark-curve}.

\bibitem{weaksub}
V.~Buterin.
\newblock Proof of {Stake}: How i learned to love weak subjectivity.
\newblock
  \url{https://blog.ethereum.org/2014/11/25/proof-stake-learned-love-weak-subjectivity/}.

\bibitem{VBopt}
V.~Buterin.
\newblock Fast verification of multiple {BLS} signatures, 2019.
\newblock
  \url{https://ethresear.ch/t/fast-verification-of-multiple-bls-signatures/5407}.

\bibitem{CHP12}
J.~Camenisch, S.~Hohenberger, and M.~{\O}. Pedersen.
\newblock Batch verification of short signatures.
\newblock 2012.

\bibitem{DecMensCon018}
A.~Decan, T.~Mens, and E.~Constantinou.
\newblock On the impact of security vulnerabilities in the npm package
  dependency network.
\newblock 2018.
\newblock \url{https://dl.acm.org/doi/10.1145/3196398.3196401}.

\bibitem{PragAg}
J.~Drake.
\newblock Pragmatic signature aggregation with {BLS}, 2018.
\newblock
  \url{https://ethresear.ch/t/pragmatic-signature-aggregation-with-bls/2105}.

\bibitem{H2CIE}
A.~Faz-Hernández, S.~Scott, N.~Sullivan, R.~Wahby, and C.~Wood.
\newblock Hashing to elliptic curves.
\newblock
  \url{https://datatracker.ietf.org/doc/draft-irtf-cfrg-hash-to-curve/}.

\bibitem{SDWAN}
S.~Gordeychik, D.~Kolegov, and A.~Nikolaev.
\newblock {SD-WAN} internet census, 2018.
\newblock \url{https://arxiv.org/pdf/1808.09027.pdf}.

\bibitem{GRADLE}
{Gradle documentation}.
\newblock Listing dependencies.
\newblock
  \url{https://docs.gradle.org/current/userguide/viewing_debugging_dependencies.html#sec:listing_dependencies}.

\bibitem{slashdec}
M.~Kalinin.
\newblock Detecting slashing conditions.
\newblock \url{https://hackmd.io/@n0ble/By897a5sH}.

\bibitem{ake}
H.~Krawczyk.
\newblock What are key exchange protocols?
\newblock
  \url{https://cyber.biu.ac.il/wp-content/uploads/2018/07/KE1\_Hugo\_BIU\_Feb2018-online.pdf}.

\bibitem{NCCblst}
NCC.
\newblock Public report – {BLST} cryptographic implementation review.
\newblock
  \url{https://research.nccgroup.com/2021/01/20/public-report-blst-cryptographic-implementation-review/}.

\bibitem{NCCBLS}
NCC.
\newblock Zcash {Overwinter} consensus and {Sapling} cryptography review.
\newblock \url{https://www.nccgroup.trust/globalassets/our-research/us/
  public-reports/2019/ncc_group_zcash2018_public_report_2019-01-30_v1.3.pdf}.

\bibitem{OhmPS020}
M.~Ohm, H.~Plate, A.~Sykosch, and M.~Meier.
\newblock Backstabber's knife collection: {A} review of open source software
  supply chain attacks.
\newblock In {\em {DIMVA} 2020}, 2020.

\bibitem{owaspapi}
OWASP.
\newblock {API} security project.
\newblock \url{https://owasp.org/www-project-api-security/}.

\bibitem{owaspasvs}
OWASP.
\newblock {Application Security Verification Standard}.
\newblock
  \url{https://owasp.org/www-project-application-security-verification-standard/}.

\bibitem{PLATE18}
I.~Pashchenko, H.~Plate, S.~E. Ponta, A.~Sabetta, and F.~Massacci.
\newblock Vulnerable open source dependencies: Counting those that matter.
\newblock In {\em ACM/IEEE International Symposium on Empirical Software
  Engineering and Measurement}, 2018.

\bibitem{zero}
N.~T.~M. Quan.
\newblock 0.
\newblock \url{https://eprint.iacr.org/2021/323}.

\bibitem{QsPrysm}
Quantstamp.
\newblock Prysm security assessment certificate.
\newblock
  \url{https://docs.prylabs.network/assets/files/Quantstamp_Prysm_Phase_0_Final_Report-d70b22fbd999b05e34346a2505782619.pdf}.

\bibitem{QsTeku}
Quantstamp.
\newblock Teku security assessment certificate.
\newblock \url{https://certificate.quantstamp.com/full/teku}.

\bibitem{rogue}
T.~Ristenpart and S.~Yilek.
\newblock The power of {Proofs-of-Possession}: Securing multiparty signatures
  against rogue-key attacks.
\newblock \url{https://eprint.iacr.org/2007/264}.

\bibitem{PFC}
Y.~Sakemi, T.~Kobayashi, T.~Saito, and R.~Wahby.
\newblock Pairing-friendly curves.
\newblock
  \url{https://datatracker.ietf.org/doc/draft-irtf-cfrg-pairing-friendly-curves/}.

\bibitem{SBCDK08}
M.~Scott, N.~Benger, M.~Charlemagne, L.~J.~D. Perez, and E.~J. Kachisa.
\newblock Fast hashing to {G2} on pairing friendly curves.
\newblock \url{https://eprint.iacr.org/2008/530}.

\bibitem{SYNOP2021}
Synopsys.
\newblock Open source security and risk analysis report 2021.
\newblock
  \url{https://www.synopsys.com/software-integrity/resources/analyst-reports/open-source-security-risk-analysis.html}.

\bibitem{TaylorVDCR20}
M.~Taylor, R.~K. Vaidya, D.~Davidson, L.~D. Carli, and V.~Rastogi.
\newblock Defending against package typosquatting.
\newblock In {\em {NSS}}, 2020.

\bibitem{ToBprysm}
{Trails of Bits}.
\newblock Prysm security assessment.
\newblock
  \url{https://docs.prylabs.network/assets/files/Trail_of_Bits_Prysm_Phase_0_Final_Report-ff2b2307a648f6b23dea9ed119b1516f.pdf}.

\bibitem{VuPaMaPlaSa}
D.~L. Vu, I.~Pashchenko, F.~Massacci, H.~Plate, and A.~Sabetta.
\newblock Towards using source code repositories to identify software supply
  chain attacks.
\newblock In {\em CCS '20}, 2020.
\newblock \url{https://doi.org/10.1145/3372297.3420015}.

\bibitem{TyCom}
D.-L. Vu, I.~Pashchenko, F.~Massacci, H.~Plate, and A.~Sabetta.
\newblock Typosquatting and combosquatting attacks on the {Python} ecosystem.
\newblock In {\em IEEE European Symposium on Security and Privacy Workshops
  (EuroS PW)}, 2020.
\newblock \url{https://ieeexplore.ieee.org/document/9229803}.

\bibitem{GOSSIPSUB}
D.~Vyzovitis, Y.~Napora, D.~McCormick, D.~Dias, and Y.~Psaras.
\newblock {GossipSub}: Attack-resilient message propagation in the {Filecoin}
  and {ETH2.0} networks.
\newblock 2020.
\newblock \url{https://arxiv.org/abs/2007.02754}.

\bibitem{WB19}
R.~S. Wahby and D.~Boneh.
\newblock Fast and simple constant-time hashing to the {BLS12-381} elliptic
  curve.
\newblock \url{https://eprint.iacr.org/2019/403}.

\bibitem{HC}
Q.~Wu and K.~Lu.
\newblock On the feasibility of stealthily introducing vulnerabilities in
  open-source software via hypocrite commits.
\newblock \url{https://linuxreviews.org/images/d/d9/OpenSourceInsecurity.pdf}.

\end{thebibliography}

\appendix

\section{Fingerprinting Queries}\label{ap:fingerprints}

 The target Ethereum consensus clients can be enumerated using popular
 search engines (Censys, FOFA, Shodan) and the queries listed below.
 The employed methodology can be defined as follows:
\begin{itemize}
  \item Research the network interfaces of the clients and identify the unique patterns that can be used for their recognition at scale. 
  \item Define and express the patterns within search engines query languages.
  \item Discover and enumerate nodes using search engines.
  \item Analyze the effectiveness of this approach for each client.
\end{itemize}
 The full methodology used for fingerprinting can be found in \cite{SDWAN}.
 For each query, we provide our estimate of the level of confidence that
 an identified host is a target host (as opposed to false positives).
This reflects the reliability of the fingeprints and queries used to identify the host and a number of possible false positives.
Note that fingerprints of some clients can't be expressed in all search
engines query language, or can be quite ineffective due to high number of false positive hosts.
In that case, we didn't provide the queries below.

\subsection{Lighthouse}

Confidence level: medium.

\paragraph{Censys.}
\begin{itemize}
    \item 443.https.get.headers.server:"Lighthouse/v1.3.0-3a24ca5/x86\_64-linux"

    \item 443.https.get.body\_sha256:"3179d38cb95e71e83a9fd64d57257c74c52c27a35c7d515f7fa256221c308b3b"
\end{itemize}

\paragraph{Shodan.}
\begin{itemize}
    \item http.html:"code" http.html:"405" http.html:"METHOD\_NOT\_ALLOWED" http.html:"stacktraces"

    \item "server: Lighthouse" "http/1.1" 405 http.html:"METHOD\_NOT\_ALLOWED"

    \item "server: Lighthouse/v1.3.0-3a24ca5/x86\_64-linux"
\end{itemize}

\subsection{Nimbus}

Confidence level: medium.

\paragraph{Shodan.}
\begin{itemize}
    \item "HTTP/1.1 411 Length Required" "Content-Length: 0" "Date" -"server" -"expires" -"cache" port:8545

    \item "HTTP/1.1 411 Length Required" "Content-Length: 0" "Date" -"server" -"expires" -"cache" port:9190
\end{itemize}

\subsection{Prysm}


Confidence level: high.

\paragraph{Censys.}
\begin{itemize}
    \item 443.https.get.title:"PrysmWebUi"
\end{itemize}

\paragraph{FOFA.}
\begin{itemize}
    \item "prysmwebui"
\end{itemize}

\paragraph{Shodan.}
\begin{itemize}
    \item http.favicon.hash:1426715472

    \item http.title:"PrysmWebUi"

    \item http.component:"tailwindcss" http.component:"google font api" all:"prysmwebui"

    \item "Content-Length: 917" all:"eth"
\end{itemize}

\subsection{Teku}

Confidence level: medium.

\paragraph{FOFA.}
\begin{itemize}
\item header="Content-Length: 131" \&\& header="Server: Javalin" \&\& port="5051"
\end{itemize}

\paragraph{Shodan.}
\begin{itemize}
\item "HTTP/1.1 404 Not Found" "Server: Javalin" "Content-Length: 131" "Content-Type: application/json" "Date"
\end{itemize}

\section{Scripts}\label{ap:scripts}

This section lists the scripts we used to find information about the
beacon clients projects and their dependencies.  
A large part of these scripts is generally applicable to any project in
the same language.

\subsection{Lighthouse (Rust)}\label{ap:scripts:rust}

{
\footnotesize
\begin{minted}{bash}
#!/bin/bash

##### Calculating max degree of dependencies
# on macOS, make sure to use GNU sed (gsed)
# the --prefix option prints the depth value before each line
cargo tree --prefix depth | grep -v '^\s*$' | sed -e 's/[0-9]\+/ & /g' -e 's/^ \| $//' 
| cut -d " " -f 1 | sort -u > depth.txt
echo -e "Dependencies max degree: $(cat depth.txt | wc -l)";

##### Count all dependencies
cargo tree --prefix none | grep -v '^\s*$' | grep  -v "(*)" | sort -u > all_dependencies.txt
sed -E 's/v[0-9]+\.[0-9]+.*$/ /g' all_dependencies.txt | sort -u > unversioned_all.txt
echo -e "Total dependencies: $(cat all_dependencies.txt | wc -l)";
echo -e "Total dependencies without version: 
$(cat unversioned_all.txt | wc -l)";

##### Count direct dependencies
rm all.txt > /dev/null 2>&1
find . -name Cargo.toml -exec cat {} + >> all.txt
for file in all.txt
do
        sed ':a;N;/\n\[.*\]/!s/\n/,/;ta;P;D' $file 
        | grep '\[dependencies]' | sed 's/,/\n/g' 
        | tail -n +2 >> direct
done
cat direct | sed '/dependencies]/d' | grep -v "path =" | cut -d " " -f 1 | awk 'NF' | grep -v '"' 
| sort -u > direct_dependencies.txt
rm direct
echo -e "Direct dependencies: $(cat direct_dependencies.txt | wc -l)";

##### Generate dependency graph
cargo depgraph > graph.dot
cat graph.dot | dot -Tpng > graph.png
echo -e "Dependency graph of the project was created at the ./graph.png file";
\end{minted}
}

\subsection{Prysm (Go)}\label{ap:scripts:prysm}


This script uses the \url{https://github.com/KyleBanks/depth} tool to help in visualizing the depth of dependencies while printing a tree with alignments depending on the degree. We could not run the tool against the Prysm, so we ran it for each of its dependencies that were GitHub projects. 
By leveraging the different alignments of the outputs, we found the max
degree. Although it does not include all the projects, it is a good estimation as the majority of modules are GitHub projects. 

{
\footnotesize
\begin{minted}{bash}
#!/bin/bash

##### Count total dependencies 
go list -m -u all | sed '1d' > all_dependencies.txt;
echo -e "Total dependencies: $(cat all_dependencies.txt | wc -l)";

##### Count direct dependencies
awk "/require/{y=1;next}y" go.mod | grep -v "indirect" | sed -n '/)/q;p' > direct_dependencies.txt;
echo -e "Direct dependencies: $(cat direct_dependencies.txt | wc -l)";

##### Count outdated dependencies
go list -m -u -json all | go-mod-outdated -update -style markdown | sed '1,2d'> outdated.txt
echo -e "Total outdated dependencies: $(cat outdated.txt | wc -l)";

##### Count outdated direct dependencies 
go list -m -u -json all | go-mod-outdated -update -direct -style markdown | sed '1,2d'> dir_outdated.txt
echo -e "Direct outdated dependencies: $(cat dir_outdated.txt | wc -l)";

##### Calculating max degree of dependencies 
cat all_dependencies.txt | grep github | cut -d " " -f 1 > all_edit.txt
cat all_edit.txt | while read line
do
        go run ~/go/pkg/mod/github.com/\!kyle\!banks/depth@v1.2.1/cmd/depth/depth.go $line >> tem
done

awk -F'[^ ]' '{print length($1),NR}' tem > tem2
var1=$(cat tem2 | sort  -nr | head -1 | cut -d " " -f 1)
# max degree is equal to the max  gap spaces /2 because there are 2 gaps for each gap, 
#+1 because I run the script against the dependencies and not the lighthouse project
var2=$(expr $var1 / 2 + 1)
# estimation because I can run it only for some GitHub dependencies and not everything
echo -e "Best estimation for dependencies max degree:"; 
echo $var2
\end{minted}
}

\subsection{Nimbus (Nim)}\label{ap:scripts:nim}

The following one-liner lists all the dependencies of a \code{.nimble}
file, the name of which is given as an argument:

{
\footnotesize
\begin{minted}{bash}
#!/bin/sh

perl -p -e 's/,\n/,/' $1 | grep requires | cut -d' ' -f2- | sed 's/ //g' | sed 's/,/\n/g' 
\end{minted}
}

\subsection{Teku (Java)}\label{ap:scripts:teku}

The following script is suitable for projects using Gradle as a package
manager, and extracts dependencies with the configuration
\code{implementation} from all the \code{build.gradle} files included in
the project.

{
\footnotesize
\begin{minted}{bash}
#!/bin/bash

##### Count direct dependencies 
find . -name "build.gradle" -exec cat {} \; | grep "implementation" | grep -v "project(" | tr -s ' ' >output 
sort -u output > direct_dependencies.txt;
rm output
echo -e "Direct dependencies:  $(cat direct_dependencies.txt | wc -l)";

##### Count all dependencies 
gradle -q dependencies | grep "\---" | grep -v "project " | cut -d "-" -f 4-7 | grep -v "(*)" | sed 's/->.*$/ /p' 
        | sed -E 's/:[0-9]+\.[0-9]+.*$/ /g' | sort -u > all_dependencies.txt
echo -e "All dependencies:  $(cat all_dependencies.txt | wc -l)";

##### Calculating max degree of dependencies 
gradle -q dependencies | sed 's/|/ /g' testfile | awk -F'[^ ]' '{print length($1),NR}' > depth.txt
# max degree is equal to the max  gap spaces /5 (because there are 5 gaps for each degree) 
# + 1 (as the 1st degree there is no gap)
echo -e "Dependencies max degree: $(expr $(cat depth.txt | sort  -nr | head -1 | cut -d " " -f 1) / 5 + 1)";
\end{minted}
}

\subsection{GitHub}\label{ap:scripts:git}

This Python script uses the GitHub REST API to extract useful project
metadata: language, number of stars, number of open issues, and date of
hte last commit.

{
\footnotesize
\begin{minted}{bash}
import sys
import requests
import argparse
import json
from termcolor import colored, cprint

parser = argparse.ArgumentParser(description='Beacon Client Dependencies scanner')
parser.add_argument('client', metavar='repo/project_name', type=str,
                    help='a beacon client GitHub repository for scanning')
parser.add_argument('file', metavar='filename', type=str,
                    help='output json file')
args = parser.parse_args()
client = args.client                       
fi = args.file
print(colored("Scanning... " + client, 'blue'))
try:
        response = requests.get("https://api.github.com/repos/"+client, timeout=5)
        response.raise_for_status()

        date = response.headers["date"]     
        print(colored("Results, as of date: " + date, "magenta"))

        dictionary = response.json()
        name = dictionary['name']
        print(colored("Beacon client: " + name, "blue"))

        language = dictionary['language']
        print(colored("Written in " + language, "magenta"))

        stars = dictionary['stargazers_count']
        print(colored("It has " + str(stars) + " stars", "magenta"))

        open_issues = dictionary['open_issues_count']
        print(colored("It has " + str(open_issues) + "open issues", "magenta"))

        last_update = dictionary['updated_at']
        print(colored("Last commit at " + last_update, "magenta" ))

        f = open(fi+".json", "w")
        f.write(json.dumps(response.json(), indent=4))
        f.close()
        print("Response written in the " + fi + ".json file")

except requests.exceptions.HTTPError as httperror:
        print(httperror)
except requests.exceptions.ConnectionError as connection:
        print(connection)
except requests.exceptions.Timeout as timeout:
        print(timeout)
except requests.exceptions.RequestException as requestexception:
        print(requestexception)
\end{minted}
}

\end{document}